\documentclass[preprint,12pt]{elsarticle}
\usepackage{amsmath}
\usepackage{amsfonts}
\usepackage{tikz-cd}
\usepackage{array}
\usepackage{enumerate}
\usepackage{mathrsfs}
\usepackage{graphicx}
\usepackage{amssymb}
\usepackage{amsthm}
\usepackage[version=4]{mhchem}
\usepackage{xcolor}
\usepackage{chemarr}
\usepackage{mathtools}
\usepackage[margin=1in]{geometry}
\usepackage{centernot}
\newtheorem{proposition}{Proposition}

\newtheorem{remark}{Remark}
\newtheorem{example}{Example}

\newcommand\myeq{\mathrel{\stackrel{\makebox[0pt]{\mbox{\normalfont\tiny$\mathcal{D}$}}}{=}}}

\newcommand{\norm}[2][\relax]{\ifx#1\relax \ensuremath{\left\Vert#2\right\Vert} \else \ensuremath{\left\Vert#2\right\Vert_{#1}}\fi}
\journal{arXiv}

\begin{document}

\begin{frontmatter}

\title{On the reduction of stochastic chemical reaction networks}

\author[label1]{Justin Eilertsen}
\author[label2,label3]{Wylie Stroberg\corref{cor1}}
\address[label1]{Mathematical Reviews, American Mathematical Society, 416 $4th$ Street, 
Ann Arbor, MI 48103, USA\\
            e-mail: {\tt jse@ams.org}}
\address[label2]{Department of Mechanical Engineering, University of Alberta, Edmonton, Alberta, Canada\\
email: {{\tt stroberg@ualberta.ca}}}
\address[label3]{Department of Biomedical Engineering, University of Alberta, Edmonton, Alberta, Canada}

\cortext[cor1]{Corresponding author}

%\date{}

%\maketitle
\begin{abstract}
The linear noise approximation (LNA) describes the random fluctuations from the mean-field concentrations of a chemical reaction network due to intrinsic noise. It is also used as a test probe to determine the accuracy of reduced formulations of the chemical master equation and to understand the relationship between timescale disparity and model reduction in stochastic environments. Although several reduced LNAs have been proposed, they have not been placed into a general theory concerning the accuracy of reduced LNAs derived from center manifold and singular perturbation theory. This has made it difficult to understand why certain reductions of the master or Langevin equations fail or succeed. In this work, we develop a deeper understanding of slow manifold projection in the linear noise regime by answering a straightforward but open question: In the presence of eigenvalue disparity, does the appropriate oblique projection of the LNA onto the slow eigenspace accurately approximate the first and second moments of complete LNA, and if not, why? Although most studies concentrate on the role of eigenvalue disparity arising from the drift matrix, we go further and examine the interplay between disparate ``drift" eigenvalues and the eigenvalues of the diffusion matrix, the latter of which may or may not be disparate. Furthermore, we place the previously established reductions of the LNA into a more general framework and formulate the necessary and sufficient conditions for the projected LNA to accurately approximate the first and second moments of the complete LNA. 
\end{abstract}

\end{frontmatter}

%\section{The reduction of stochastic reaction network models: Bridging the gap from continuous to discrete state space}
\section{Introduction}

The derivation of accurate reduced models of chemical reactions is a coveted element of mathematical and computational biology. Low-dimensional deterministic ordinary differential equation (ODE) models of biochemical reactions play a critical role in drug design and drug targeting: kinetic parameters are estimated by fitting experimental timecourse data to reduced ODE models such as the standard Michaelis-Menten rate law~\cite{Stroberg2016}. In addition, the reduction of stochastic models permits a favorable trade-off between accuracy and computational complexity: a high-dimensional stochastic model of a biochemical reaction can often be replaced with a low-dimensional model with a negligible cost in accuracy and a substantial reduction in computational complexity~\cite{Cao,Sanft}. 

The mathematical feature that permits model reduction is \textit{timescale separation}. If a reaction mechanism is comprised of several elementary reactions, timescale separation implies that the corresponding rates of the elementary reactions are disparate: the rates of a subset of elementary reactions are substantially -- and consistently -- less than the rates of the remaining elementary reactions throughout the reaction's timecourse~\cite{OthmerI,OthmerII}.

The challenge in reducing a chemical reaction network partially lies in how reactions are modeled: Depending on the size of the system, a reaction may be modeled deterministically or stochastically. Moreover, stochastic models come in several varieties, ranging from the chemical master equation (CME), which represents the reaction as a continuous-time, discrete state-space Markov process~\cite{GILLESPIECME}, to the chemical Langevin equation (CLE), which is a nonlinear stochastic differential equation (SDE) model driven by multiplicative noise~\cite{kurtz1978}, and finally, the linear noise approximation, which is a linear SDE model driven by additive noise. The choice of model depends on several factors including the size and spatial homogeneity of the system.

In the thermodynamic limit and in the absence of diffusion, reaction networks are commonly modeled according to the law of mass action, in which case the temporal evolution of each species' concentration obeys a deterministic ordinary differential equation. In this context, timescale separation is synonymous with eigenvalue disparity, and model reduction is achievable through the application of center manifold theory~\cite{Roberts2015,Roberts1996,Roberts2013,KnoblochCM} or singular perturbation theory~\cite{Berglund,Yin2004,Yin2005} (the latter, which can be viewed as a special case of the former, is also known as geometric singular perturbation theory or Tikhonov/Fenichel theory). In chemical kinetics, the low-dimensional models that result from the application of singular perturbation theory are called {\it quasi-steady-state approximations} (QSSAs). Unfortunately, the aforementioned deterministic reduction methods do not always apply in a straightforward way to stochastic models. Even when they do, they often require the system to be expressed in ``special coordinates" that separate the system into distinct fast and slow processes. This is restrictive for two reasons: First, singular perturbation theory is coordinate independent, so there is no need to perform a coordinate transformation in order to reduce deterministic ODE models. Second, even if there is a tractable coordinate transformation that allows the deterministic model to be expressed in fast and slow coordinates, these new coordinates may not be experimentally measurable or even chemically meaningful. Moreover, given Ito's lemma, nonlinear coordinate transformations require special care in the CLE regime.

Due to the inherent difficulty of model reduction in various stochastic regimes, there is a large body of literature probing the accuracy of so-called \textit{heuristically} reduced stochastic models. In the heuristic approach, stochastic models are adapted from deterministic quasi-steady-state approximations, but the justification for the adaptation is not necessarily rigorous. Thus, the accuracy of heuristic reductions is doubtful, and the mixed reviews published in the literature reflect this. An example from biochemistry is the stochastic QSSA to the CME of the Michaelis-Menten reaction mechanism, first introduced by~\citet{Rao2003}. Using the total substrate introduced by~\citet{Borghans1996}, Arkin and Rao~\cite{Rao2003} reported the stochastic QSSA accurately approximates the mean and variance of the total substrate and concluded that timescale separation was sufficient to ensure the accuracy of the heuristic reduction. In other words,~\citet{Rao2003} concluded that the same conditions required to ensure the accuracy of the deterministic reduction also ensure the accuracy of the stochastic QSSA. Later studies provided a more rigorous justification for the stochastic QSSA. By directly reducing the CME,~\citet{Mastny2007} found that the stochastic QSSA is accurate at very high and very low free enzyme concentrations, but the authors did not address the accuracy of the stochastic QSSA at intermediate free enzyme concentrations. A later study by \citet{Sanft} came to the same conclusion as~\citet{Rao2003}, but carefully noted that the stochastic QSSA can {\it overestimate} the variance in certain cases. 

However, several other authors came to a different conclusion.~\citet{GrimaBreakdown} first reported on the breakdown of the stochastic QSSA in the presence of intrinsic noise. Later, by utilizing the strategy outlined by~\citet{Janssen1989} involving the LNA, \citet{Thomas2011} demonstrated conclusively that timescale separation is necessary -- but not sufficient -- to ensure the accuracy of the stochastic QSSA at intermediate free enzyme concentration. A later study conducted by~\citet{KIM2014} arrived at the same conclusion: timescale separation alone (eigenvalue disparity) does not justify the stochastic QSSA. 

In essence, a broad review of the literature on heuristic reduction ultimately concludes that sometimes heuristic reductions merely require timescale separation to ensure accuracy, but sometimes they require ``more." The puzzling question is not what the ``more" is, as this can often be determined through brute-force calculations~\cite{Agarwal2012,Thomas2011}, but {\it why} these additional qualifiers are required to ensure the validity of certain stochastic reduced models? 

Although the ultimate goal is to understand the technical subtleties of model reduction across the thermodynamic spectrum from continuous to discrete state spaces, several questions need to be answered before attempting to bridge such a large gap. Many of the more conclusive (and rigorous) analyses focus on the LNA, primarily because of its linearity, which makes its analysis comparatively more straightforward than that of the CLE or CME. Moreover, it is well-known that, for zeroth- and first-order reaction networks, the first and second moments of the LNA and the CME are in agreement~\cite{BurrageCLE}, and sometimes this agreement extends to second order reaction networks~\cite{GrimaSecondOrder}. However, in order to understand the relationship between timescale separation and model reduction in the LNA regime, {\it we must understand how to systematically reduce the LNA}, and this is where the literature seems to be putting the ``cart before the horse." To the best of our knowledge, there have been no definitive analyses addressing the coordinate-independent reduction of the LNA. Although several accurate reductions of the LNA have been reported, beginning with~\citet{PAHLAJANI201196}, and later with the {\it slow scale} LNA or ``ssLNA" derived by~\citet{Thomas2012}, neither are coordinate independent and apply only to a small subset of singularly perturbed reaction networks (i.e., you cannot necessarily apply the reduction strategies of~\citet{PAHLAJANI201196} or~\citet{Thomas2012} to more general singularly perturbed reaction networks). 

In this paper, we ask a relatively simple question: Given that the mean-field approximation determined by mass action kinetics rests at an attracting, stationary node, does the oblique projection of the LNA onto the slow eigenspace of the Jacobian (along a direction parallel to the fast eigenspace) provide an accurate reduction of the LNA and, if not, why? Remarkably, this simple -- and fundamental -- question has not been addressed in previous studies of chemical reaction networks. If the goal is to ultimately understand why certain reduction techniques succeed or fail when applied or adapted to specific stochastic environments, it is essential to first understand when the most basic reduction technique (slow eigenspace projection) provides a reliable and accurate reduction of the LNA. Moreover, this question directly relates to the question surrounding the necessity and sufficiency of timescale separation, since the presence of a spectral gap (disparate eigenvalues) implies the existence of fast and slow eigenspaces. 

The paper is summarized as follows. We recall and define the linear noise approximation in Section 2 and formulate the central aim of our paper in mathematical terms. In Section 3 we briefly review the components of deterministic singular perturbation theory required for the analysis. In Section 4 we derive necessary conditions that ensure that the reduced LNA (obtained from slow eigenspace projection) converges to the long-time mean and covariance of the full LNA. In Section 5, we take a detailed look at several examples and explain why some of the reduced models presented in the literature, such as the slow-scale linear noise approximation of~\citet{Thomas2012} and the total quasi-steady-state approximation championed by~\citet{KIM2015} are so accurate. In Section 6, we conclude with a discussion of the role timescale separation plays in the context of stochastic model reduction applied to chemical reaction networks, provide an overview of the results obtained from our analysis, and suggest possible avenues for future work.

\section{The linear noise approximation}

In the deterministic limit and in the absence of diffusion, chemical reaction networks are modeled by mass-action kinetics. If the network consists of ``$n$" chemical species and ``$k$" elementary reactions, the temporal evolution of the concentration of each species, $x_i$, is determined by the system of ordinary differential equations,
\begin{equation}\label{CRN}
\dot{x}_i = \sum_{j=1}^k S_{ij}r_j(x)=: f_i(x), \quad 1 \leq i \leq n
\end{equation}
where ``$\dot{\phantom{x}}$" denotes differentiation with respect to time, $t$, $r_j(x)$ is the rate of the ``$jth$" elementary reaction, and $S_{ij}\in \mathbb{Z}^{n\times k}$ is a (net) stoichiomentric matrix.

If the system \eqref{CRN} has a unique, stable fixed point $x=x^*$, then, after a transient phase, the presence of intrinsic noise will precipitate random fluctuations about $x=x^*$. As long as the size of the system, $\Omega$, is adequately large, the linear noise approximation says that the fluctuations, $Y$, satisfy the Ornstein-Uhlenbeck process
\begin{equation}
{\rm d}Y_i = \sum_{j=1}^n A_{ij}Y_j\;{\rm d}t + \gamma\sum_{m =1}^k B(x^*)_{im}\;{\rm d}W_m(t),\quad \gamma = \Omega^{-1/2},
\end{equation}
where the drift, $A_{ij}$, and diffusion, $B_{im}$, terms are given by
\begin{equation}
A_{ij} = \cfrac{\partial f_i(x)}{\partial x_j}\bigg|_{x=x^*}, \qquad B_{im}(x^*) = S_{im}\sqrt{r_m(x^*)},
\end{equation}
and $W_m(t)$ are standard Brownian motions:
\begin{subequations}
\begin{align}
\mathbb{E}\{W(t)\} &= 0\\
\mathbb{E}\{W(t)W(s)\} &= \min\{t,s\}. 
\end{align}
\end{subequations}

Our interest is in singularly perturbed LNAs of the form
\begin{equation}\label{SDE0}
{\rm d}Y = (A_0 + \varepsilon A_1) Y \;{\rm d}t + \gamma \cdot B(x^*;\sqrt{\varepsilon})\; {\rm d}W(t).
\end{equation}
In the limit $(\varepsilon,\gamma)\to (0,0)$, the SDE \eqref{SDE0} reduces to the linear ODE
\begin{equation}\label{linD}
\dot{Y}=A_0Y,
\end{equation}
where the $A_0 \in \mathbb{R}^{n \times n}$ is singular. Central to our analysis will be the assumption that eigenspectrum of $A_0$ is comprised of a zero eigenvalue with an algebraic and geometric multiplicity, $r$, with the remaining $n-r$ eigenvalues real and strictly negative. Under this assumption, $\mathbb{R}^n$ admits the splitting
\begin{equation*}
\mathbb{R}^n = E^- \oplus E^0
\end{equation*}
where $E^-$ is the $(n-r)$-dimensional fast eigenspace of $A_0$, and $E^0$ is the corresponding $r$-dimensional center subspace. In this situation, the center subspace $E^0$ constitutes a normally hyperbolic and invariant manifold. In Section 3 we recall some basic facts from deterministic theory, but for now it suffices to say that the ``deterministic" approach to reduction is to simply project \eqref{SDE0} onto $E^0$,
\begin{equation}\label{SDE0P}
{\rm d}Y = \pi_0 \varepsilon A_1 Y \bigg|_{Y\in E^0}\;{\rm d}t + \gamma \cdot \pi_0 B(x^*;\sqrt{\varepsilon})\; {\rm d}W(t),
\end{equation}
where $\pi_0$ is the unique projection matrix that projects $v\in \mathbb{R}^n$ onto $E^0$:
\begin{subequations}
\begin{align}
\pi_0 &: \mathbb{R}^n \to E^0,\\
I-\pi_0&:\mathbb{R}^n \to E^-,
\end{align}
\end{subequations}
with $I$ denoting the $n\times n$ identity matrix. Formally, the specific question we address in this paper is under what circumstances do the {\it steady-state} first and second moments (mean and covariance) of the projected LNA \eqref{SDE0P} converge to the steady-state mean and covariance of the complete LNA \eqref{SDE0} as $\varepsilon \to 0$?

Several observations are worth mentioning. First, $\pi_0$ is an oblique projection operator, and therefore one should not assume that $\pi_0 = \pi_0^T$ (in almost all cases this will not be true). Second,  we have chosen to set $\gamma =1$. This is because, while the size of the system
determines the intensity of the noise, it is not central to our analysis and plays a somewhat inert role. Third, we will generally denote the components of $Y$ in lowercase with integer subscripts $y_i$ or subscripts that indicate the fluctuations pertain to a specific chemical species concentration (i.e., $y_c$ would denote the fluctuations in the concentration of species ``$c$").

Finally, since our interest is on the first and second moments, we will decompose the analysis of \eqref{SDE0P} into two parts: the deterministic evolution of the mean, $\mathbb{E}\{Y\}$,
\begin{equation}\label{mean}
\cfrac{d\mathbb{E}\{Y\}}{dt} = A\mathbb{E}\{Y\},
\end{equation}
and the covariance, $Z\in \mathbb{R}^{n\times n}$,
\begin{equation}\label{cov}
\cfrac{dZ}{dt} = AZ+ZA^T + BB^T
\end{equation}
where $A=A_0+\varepsilon A_1$ and $B= S\sqrt{{\rm diag} \;(r(x^*))}$. Both \eqref{mean} and \eqref{cov} are ordinary differential equations and therefore singular perturbation methods are directly applicable. 

\section{Brief review of geometric singular perturbation theory}

Before starting the analysis of the projected LNA \eqref{SDE0P}, it will help to review some basic facts from geometric singular perturbation theory (GSPT) as they apply to linear autonomous differential equations; further details can be found in~\cite{Fenichel1979,Wechselberger2020,kuehn2015,HekGSPT}. For simplicity, consider the two-dimensional linear system,
\begin{equation}\label{lin}
\dot{x} = (A_0 + \varepsilon A_1)x, \quad x(0) = x_0,
\end{equation}
and assume that the origin is a stable node and therefore the eigenspectrum of $A$ consists of two distinct and strictly negative eigenvalues: the fast eigenvalue, $\lambda_-$, and the slow eigenvalue, $\lambda_+$, both of which are analytic with respect to $\varepsilon$ and admit expansion(s)\footnote{See Appendix for details regarding this assumption.}
\begin{subequations}
\begin{align}
\lambda_- &= \lambda_-^{(0)} + \varepsilon \lambda_{-}^{(1)}+ \mathcal{O}(\varepsilon^2)\\
\lambda_+ &= \lambda_+^{(0)} + \varepsilon \lambda_{+}^{(1)}+ \mathcal{O}(\varepsilon^2) =  \varepsilon \lambda_{+}^{(1)}+ \mathcal{O}(\varepsilon^2).
\end{align}
\end{subequations}

Setting $\varepsilon =0$ in \eqref{lin} results in what is known as the {\it layer problem} or {\it fast subsystem}. Under the assumption that the eigenspectrum of $A_0$ consists of one strictly negative eigenvalue, $\lambda_-^{(0)}$, and one zero eigenvalue, $\lambda_+^{(0)} =0$, the one-dimensional center subspace, $E^0$, consists entirely of equilibrium solutions with respect to the \textit{layer problem}. Moreover, $E^0$ is normally hyperbolic and, therefore, persists, along with its stable ($W^s(E^0)$) and unstable ($W^u(E^0))$ manifolds, whenever $\varepsilon $ is nonzero but sufficiently small. In essence, normally hyperbolic manifolds can be thought of as the higher-dimensional analogue of a hyperbolic fixed point: they are structurally stable with respect to smooth perturbations. This is important because the normal hyperbolicity of $E^0$ ensures that when the perturbation is activated (that is, $\varepsilon >0$ and sufficiently small), $\mathbb{R}^2$ will continue to contain a normally hyperbolic and invariant manifold. For linear systems of the form \eqref{lin}, the slow manifold is the slow eigenspace $E^s$, of $A=(A_0+\varepsilon A_1)$. 

Once the perturbation is turned on (assuming the origin is a stable node), the solution trajectories (integral curves) will rapidly move towards $E^s$ and continue to approach the origin parallel to the direction of $E^s$ as $t\to \infty$. And, while the linear system \eqref{lin} has a well-known solution
\begin{equation}\label{exact}
x(t) = \exp(t(A_0+\varepsilon A_1))x_0,
\end{equation}
our interest will be in constructing approximate solutions that converge to solutions of \eqref{lin} as $\varepsilon \to 0.$ The motivation here is that we are ultimately interested in reducing linear and time-invariant SDEs that model reaction networks, and therefore we want to ``separate" the fast and slow timescale contributions to the exact solution in order to clearly understand why such a procedure may fail to produce a reliable reduced model in the linear noise regime. To do this, we apply Fenichel theory~\cite{Fenichel1979,Wechselberger2020} and project \eqref{lin} onto $E^0$. The matrix
\begin{equation}
 \pi_0 :=   I - (\lambda_-^{(0)})^{-1}A_0: \mathbb{R}^2 \to E^0
\end{equation}
projects onto $E^0$ along $E^-$ (the fast eigenspace of $A_0$). A straightforward projection yields
\begin{equation}\label{red}
\dot{x} = \varepsilon \pi_0A_1 x, \quad x\in E^0,
\end{equation}
which, again, is referred to as a quasi-steady-state approximation in chemical kinetics~\cite{Kumar1998,Wilhelm2000}. Notice that we can express \eqref{red} in terms of a slow time, $\tau = \varepsilon t$,
\begin{equation*}
x' = \pi_0A_1x.
\end{equation*}
where ``$\phantom{x}'$" denotes differentiation with respect to slow time, $\tau$.

\begin{figure}[htb!]
\centering
\includegraphics[width=10.0cm]{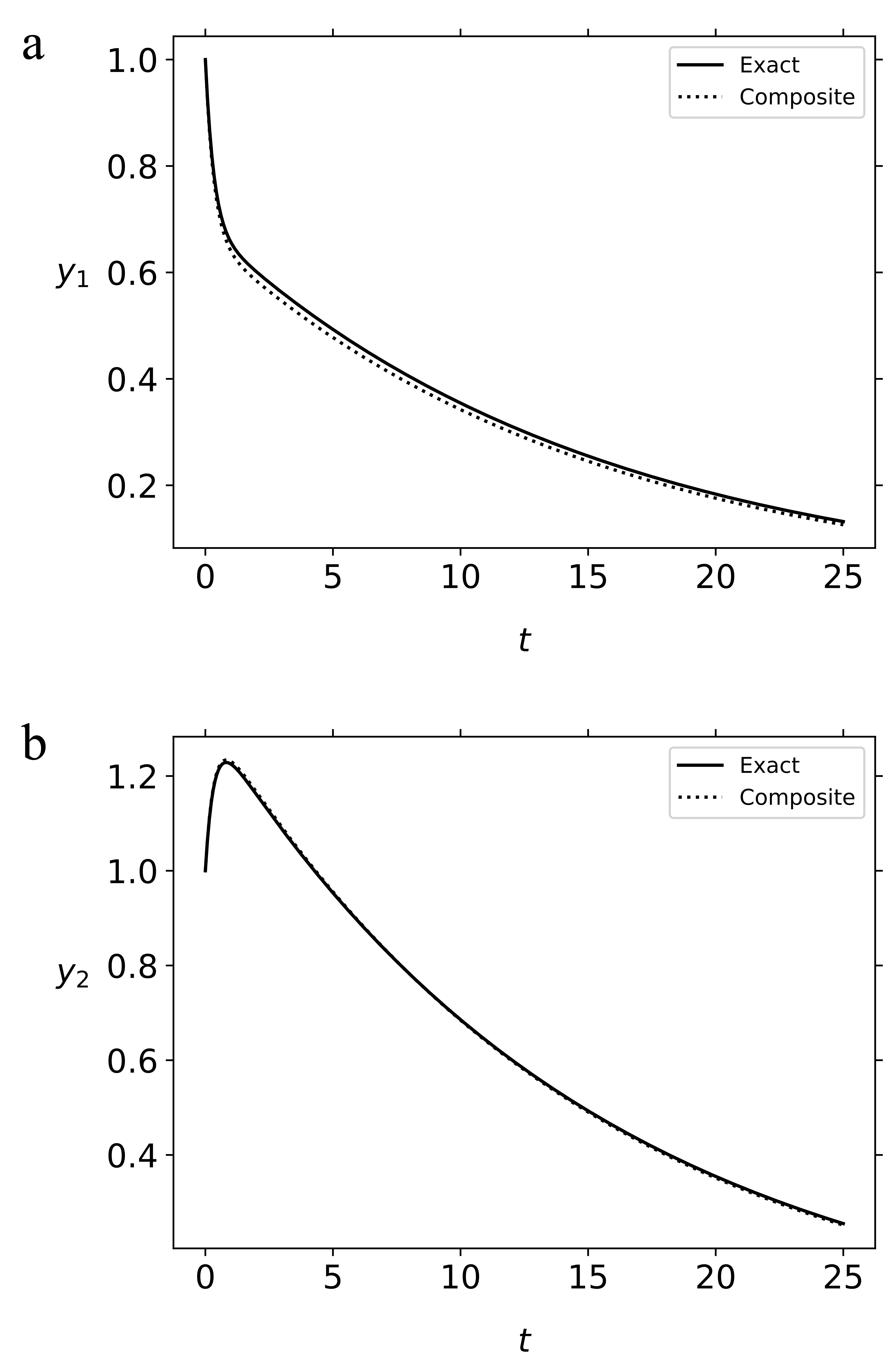}
\caption{\textbf{The composite expansion \eqref{composite} approximates the exact solution \eqref{exact} over fast and slow timescales.} In this example, $A_0=\begin{pmatrix}-2&\;\;\;1\\\;\;\;2&-1\end{pmatrix}$, $A_1=\begin{pmatrix}0&\;\;\;0\\0&-\varepsilon \end{pmatrix}$, and $Y(0)=\begin{pmatrix}1\\1\end{pmatrix}$. A simple calculation reveals $\pi_0=\begin{pmatrix}1/3 & 1/3 \\ 2/3 & 2/3\end{pmatrix}$, which yields $y_1(t,\tau) \approx (1/3)\exp(-3t) + (2/3)\exp(-(2/3)\tau)$ and $y_2(t,\tau)\approx -(1/3)\exp(-3t) + (4/3)\exp(-(2/3)\tau)$. The solid curves are the numerical solutions to the components of full system \eqref{exact}, and the dotted curves are the solutions to the components of the composite expansion \eqref{composite} with $\varepsilon = 0.1$. It is straightforward to see that the composite expansion approximates the exact solution over the fast and slow phases of the timecourse, and improves as $\varepsilon \to 0$.} \label{FIG1}
\end{figure}

While \eqref{red} does approximate the dynamics of \eqref{lin} on the slow timescale, we cannot expect solutions to \eqref{red},
\begin{equation}\label{slowflow}
x(\tau) = \exp(\tau \pi_0 A_1)\pi_0x_0
\end{equation}
to approximate solutions to \eqref{lin} as $\varepsilon \to 0 $ unless $x_0$ sufficiently close to $E^0$ since \eqref{slowflow} approximates the flow \textit{on} the slow eigenspace of $A$, $E^s$, but does not account for the behavior of trajectories in the {\it approach} to $E^s$. To construct an approximation that holds over fast and slow timescales, we must \textit{match} the fast and slow solutions:
\begin{equation}\label{composite}
x(t,\tau) \approx \exp(tA_0)(I-\pi_0)x_0 + \exp(\tau \pi_0A_1)\pi_0x_0,
\end{equation}
which approximates the exact solution \eqref{exact} over both timescales as $\varepsilon \to 0.$ Formally, the approximation \eqref{composite} is called a \textit{composite} expansion; see {{\sc figure}} \ref{FIG1} for a numerical illustration of this method.

The utility of the composite expansion resides in the perspective it provides in understanding the behavior of the solution over fast and slow timescales. In the {\it inner} approximation, it is simply\footnote{Recall that $(I-\pi_0)x_0$ lies entirely in the image of $A_0$ which is the fast eigenspace of $A_0$.}
\begin{equation}
x^i(t) := \exp(tA_0)(I-\pi_0)x_0 + \pi_0x_0 = \exp(t\lambda_-^{(0)})(I-\pi_0)x_0 + \pi_0x_0,
\end{equation}
which describes the \textit{fast} exponential decay of the initial condition component belonging to $E^-$ while the component belonging to $E^0$ remains constant on the fast timescale, $t$. The {\it outer solution}, \eqref{red}, approximates the \textit{slow} exponential decay of the flow on the invariant slow manifold, $E^s$. The term ``matched" follows from the requirement that the limiting behavior of the fast dynamics must equal the initial condition of slow dynamics:
\begin{equation*}
    \lim_{t \to \infty} \exp(tA_0)x_0 = \pi_0x_0,
\end{equation*}
while the term {\it composite} signifies that the approximation \eqref{composite} is formed by fusing together (via subtraction of the overlapping term, $\pi_0x_0$) the approximate solutions to \eqref{lin} over the respective fast and slow timescales, $t$ and $\tau$.

Moving forward, the objective will be to reconsider the fast and slow components of the LNA \eqref{SDE0} and to build a composite expansion that approximates both the drift and the diffusion of \eqref{SDE0} on the fast and slow timescales. The rationale is that separating fast and slow contributions will allow us to better understand the multiscale drift and diffusion behavior of \eqref{SDE0} and the utility of the projected LNA \eqref{SDE0P}. 

\section{Reduction of linear, time-invariant SDEs via singular perturbation methods}

In this section, we systematically analyze the ability of the projected LNA \eqref{SDE0P} to reliably and accurately estimate the mean and variance of the complete LNA \eqref{SDE0}. We begin with the mean, or {\it expectation}, $\mathbb{E}\{Y\}.$

\subsection{Analysis of the mean}

The temporal behavior of the mean, $\mathbb{E}\{Y\}$, admits an exact solution,
\begin{equation}
\mathbb{E}\{Y\}(t) = \exp(tA)\mathbb{E}\{Y(0)\}.
\end{equation}
However, if we wish to get a clear perspective on the behavior of the mean on the fast and slow timescales, we can once again build an asymptotic approximation following the same methodology discussed in Section 2. This yields
\begin{equation}\label{composite2}
\mathbb{E}\{Y\}(t,\tau) \approx \exp(tA_0)(I-\pi_0)\mathbb{E}\{Y(0)\} + \exp(\tau \pi_0A_1)\pi_0\mathbb{E}\{Y(0)\}.
\end{equation}

Again, if $n=2$ and $A\in \mathbb{R}^{2\times 2}$, the zeroth-order drift matrix, $A_0$, has two eigenvalues: the trivial eigenvalue, $\lambda_+^{(0)}=0$, and the fast eigenvalue, $\lambda_-^{(0)}.$ Moreover, it holds that\footnote{See Appendix for the details concerning this assertion.}
\begin{equation}
\pi_0A_1\pi_0\mathbb{E}\{Y(0)\} = \lambda_+^{(1)}\pi_0\mathbb{E}\{Y(0)\},
\end{equation}
and therefore \eqref{composite2} reduces to
\begin{equation}\label{composite3}
\mathbb{E}\{Y\}(t,\tau) \approx \exp(\lambda_{-}^{(0)}t)(I-\pi_0)\mathbb{E}\{Y(0)\} + \exp( \lambda_+^{(1)}\tau)\pi_0\mathbb{E}\{Y(0)\}.
\end{equation}

In conclusion, reducing the LNA via projection onto $E^0$ does not account for any drift that occurs on the fast timescale. Consequently, the projected LNA cannot approximate $\mathbb{E}\{Y\}$ unless the expectation of the initial condition, $\mathbb{E}\{Y(0)\}$, is identical to $0$ or lies sufficiently close to the center subspace. If the initial conditions lie far enough away from the slow eigenspace, then it may be necessary to use a composite expansion to approximate the mean across both timescales.

On the other hand, there is nothing problematic about the projection of the drift term onto the center subspace $E^0$ when it comes to the long-time accuracy of the expectation since $\mathbb{E}\{Y\} \to 0$ as $t\to \infty$. Thus, the point is that the discrepancies between the drift behavior of \eqref{SDE0P} and \eqref{SDE0} can usually be minimized by choosing an appropriate initial condition. If we choose the initial state of the system to be $\mathbb{E}\{Y(0)\}=0$, then any differences between \eqref{SDE0P} and \eqref{SDE0} {\it must} emerge from the projection of the diffusion terms onto $E^0$. We examine this hypothesis in Subsection \ref{cov8}.
%\begin{proposition}
%The projected LNA \eqref{SDE0P} converges to the expectation of the complete LNA \eqref{SDE0} as $\varepsilon \to 0$ for any initial condition, $\mathbb{E}\{Y(0)\}$, that lies entirely within $E^s$, the slow eigenspace of $A=A_0+\varepsilon A_1$.
%\end{proposition}
%\begin{proof}
%Let $\mathbb{E}\{Y_p\}$ denote the expectation of the projected LNA \eqref{SDE0P}. With $\mathbb{E}\{Y(0)\}\in E^s$, the expectation of the projected LNA is
%\begin{equation}\label{SLF}
%\begin{aligned}
%\mathbb{E}\{Y_p(t)\} &= \exp(\varepsilon t\pi_0A_1)\pi_0\mathbb{E}\{Y(0)\},\\
%&= \exp(\varepsilon t\lambda_+^{(1)})\pi_0\mathbb{E}\{Y(0)\}.
%\end{aligned}
%\end{equation}

%Correspondingly, the expectation of the full LNA, $\mathbb{E}\{Y\}$ with $\mathbb{E}\{Y(0)\}\in E^s$ is
%\begin{equation}
%\mathbb{E}\{Y(t)\} = \exp(t\lambda_+)\mathbb{E}\{Y(0)\}
%\end{equation}
%and, from Fenichel's theorem~\cite{Fenichel1979}, converges to \eqref{SLF} as $\varepsilon \to 0$.
%\end{proof}

\subsection{Analysis of the covariance}\label{cov8}
For a linear, time-invariant SDE driven by additive noise, the convergence of $\mathbb{E}\{Y\}$ is actually an automatic consequence of Fenichel theory. The follow-up question is whether or not a similar consequence holds for the covariance. Let $Y= (y_1\;\;y_2)^T$. The deterministic evolution of the covariance matrix, ``$Z$,"  
\begin{equation}
Z = \begin{pmatrix}\mathbb{E}\{(y_1-\mathbb{E}\{y_1\})(y_1-\mathbb{E}\{y_1\})\}&\mathbb{E}\{(y_1-\mathbb{E}\{y_1\})(y_2-\mathbb{E}\{y_2\})\}\\ \mathbb{E}\{(y_2-\mathbb{E}\{y_2\})(y_1-\mathbb{E}\{y_1\})\} & \mathbb{E}\{(y_2-\mathbb{E}\{y_2\})(y_2-\mathbb{E}\{y_2\})\}\end{pmatrix} =:\begin{pmatrix}z_{11}& z_{12}\\z_{21}&z_{22}\end{pmatrix},
\end{equation}
is determined by the Lyapunov matrix equation,
\begin{equation}\label{lyap}
\dot{Z} = \mathcal{L}(Z) + B(x^*)B(x^*)^T,
\end{equation}
where the operator $\mathcal{L}: \mathbb{R}^{2\times 2} \to \mathbb{R}^{2\times 2}$, and its adjoint, $\mathcal{L}^{\dagger}$, are
\begin{subequations}
\begin{align}
\mathcal{L}(Z):= AZ + ZA^T,\\
\mathcal{L}^{\dagger}(Z):= A^TZ + ZA.
\end{align}
\end{subequations}
If $v_{\pm}$ and $\lambda_{\pm}$ are eigenvalue/eigenvector pairs of $A$, then
\begin{equation}
\{v_-v_-^T, 2\lambda_-\}, \quad \{v_-v_+^T, \lambda_-+\lambda_+\},\quad \{v_+v_-^T, \lambda_++\lambda_-\},\quad \{v_+v_+^T, 2\lambda_+\}
\end{equation}
are the corresponding eigenvectors and eigenvalues of $\mathcal{L}.$

The specific question we will address in this subsection is whether or not the covariance obtained from the projected LNA \eqref{SDE0P}
reliably approximates the covariance of the LNA. Let us also assume that there are $``k$" elementary reactions with ``$f$" fast reactions and ``$s$" slow reactions (i.e., $k=s+f$). This allows us to express the underlying mass action equations in perturbation form
\begin{equation}
\dot{x} = S_0r_0 + \varepsilon S_1 r_1, \quad S_0\in \mathbb{R}^{2\times f},\quad r_0\in \mathbb{R}^f, \quad S_1\in \mathbb{R}^{2\times s},\quad r_1\in \mathbb{R}^s
\end{equation}
where $r_0$ is a column vector whose entries are the  $\mathcal{O}(1)$ rates of the fast reactions and $r_1$ is a column vector whose entries are the $\mathcal{O}(\varepsilon)$ rates of the slow reactions. The corresponding LNA is therefore expressible as
\begin{equation}\label{mod}
{\rm d}Y = (A_0+\varepsilon A_1)Y {\rm d}t + B_0(x^*){\rm d}W_0(t) + \sqrt{\varepsilon}B_1(x^*){\rm d}W_1(t)
\end{equation}
where ${\rm d}W_0\in \mathbb{R}^{f}$, ${\rm d}W_1\in \mathbb{R}^{s}$ and 
\begin{subequations}
\begin{align}
B_0(x^*) &= S_0\sqrt{{\rm diag}\;(r_0(x^*))},\\
B_1(x^*) &= S_1\sqrt{{\rm diag}\;(r_1(x^*))}.
\end{align}
\end{subequations}

To begin our analysis, we set $\varepsilon =0$ in \eqref{mod}, which yields
\begin{equation}\label{modf}
{\rm d}Y = A_0 Y {\rm d}t + B_0(x^*){\rm d}W_0(t).
\end{equation}
The zeroth-order LNA \eqref{modf} describes the stochastic evolution of $Y$ on the fast timescale. Thus, we will refer to \eqref{modf} as the {\it fast} LNA. The covariance of the fast LNA, $Z_f$, evolves according to
\begin{equation}
\dot{Z}_f = \mathcal{L}_0(Z_f) + B_0B_0^T,
\end{equation}
where $\mathcal{L}_0(Z_f)= A_0Z_f+Z_fA_0^T.$

Central to singular perturbation theory is the set of stationary points $\mathcal{M}_0$, for which $\dot{Z}_f=0$. This set is formally given by
\begin{equation}\label{CM}
\mathcal{M}_0 = \{X\in \mathbb{R}^{2\times 2}: \mathcal{L}_0(X) = -B_0B_0^T\}.
\end{equation}
\begin{remark}
We have introduced the notation $\mathcal{M}_0$ here as a reminder that the set of stationary points is not necessarily a vector subspace of $\mathbb{R}^{2\times 2}.$ While $\mathcal{M}_0$ is a linear manifold, it may or may not contain the zero vector, and therefore does not automatically qualify as a vector space.
\end{remark}

The linear system that defines $\mathcal{M}_0$ admits an infinite number of non-trivial solutions as long as the Fredholm alternative holds. Specifically, it must hold that
\begin{equation}\label{FAL}
w^TB_0B_0^Tw =0, \quad \forall w \;\;s.t.\;\;A_0^Tw=0.
\end{equation}
Provided \eqref{FAL} holds, the set $\mathcal{M}_0$ is given by $Z_H + Z_P$, where
\begin{equation}
Z_H = {\rm span} \{v_+^{(0)}v_+^{(0),T}\}, \quad \mathcal{L}_0(Z_P) = -B_0B_0^T,
\end{equation}
and $A_0v_+^{(0)}=0.$ An important consequence of the Fredholm alternative is the following:
\begin{proposition}\label{prop1}
If the Fredholm alternative \eqref{FAL} holds, then $\pi_0B_0=0$ and therefore the column space of $B_0$ lies entirely within the image of $A_0$. Consequently,
\begin{equation}
\mathcal{M}_0 = {\rm span}\{v_+^{(0)}v_+^{(0),T}\} + \cfrac{1}{2|\lambda_-^{(0)}|}B_0B_0^T.
\end{equation}
\end{proposition}
\begin{proof}
If the Fredholm alternative applies, then the projection of $B_0B_0^T$ onto the kernel of $\mathcal{L}_0$ along its image must vanish
\begin{equation}
\pi_0B_0B_0^T\pi_0^T = 0,
\end{equation}
which implies $\pi_0B_0=0$.
\end{proof}

Repeating the techniques employed in our analysis of the mean, the {\it inner} approximation to the covariance equation \eqref{lyap} is
\begin{equation}\label{IC}
Z_f(t)= \exp(2\lambda_-^{(0)}t)(I-\pi_0)Z(0)(I-\pi_0^T) + \cfrac{B_0B_0^T}{2|\lambda_-^{(0)}|}(1-\exp(2\lambda_-^{(0)}t)) + \pi_0Z(0)\pi_0^T.
\end{equation}
Since $\lambda_-^{(0)}<0$, the long-time solution to \eqref{IC} approaches
\begin{equation}\label{LTF}
\lim_{t\to \infty} Z_f(t) = \cfrac{B_0B_0^T}{2|\lambda_-^{(0)}|} + \pi_0Z(0)\pi_0^T.
\end{equation}
The first term on the right-hand side of \eqref{LTF} accounts for diffusion that occurs on the fast timescale. The second term emerges because the component of $Z(0)$ that belongs to $\ker \mathcal{L}_0$ is effectively ``frozen" on the fast timescale and only evolves on the slow timescale, $\tau = \varepsilon t$. 

On the slow timescale, the evolution of the covariance is approximated by the Fenichel reduction,
\begin{equation}\label{OSS}
Z_s' = \pi_0\mathcal{L}_1(Z)\bigg|_{Z\in \mathcal{M}_0} \pi_0^T+ \pi_0 B_1B_1^T\pi_0^T,
\end{equation}
where again ``$\phantom{Z}'$" denotes differentiation with respect to the slow timescale, and $\mathcal{L}_1(Z) = A_1Z+ZA_1^T$. Integrating \eqref{OSS} provides the outer solution that approximates the behavior of $\pi_0Z(0)\pi_0^T$ on the slow timescale:
\begin{equation}
Z_s(\tau) = \exp(2\lambda_+^{(1)}\tau)\pi_0Z(0)\pi_0^T + \cfrac{\pi_0B_1B_1^T\pi_0^T}{2|\lambda_+^{(1)}|}(1-\exp(2\lambda_+^{(1)}\tau)).
\end{equation}

The composite expansion to \eqref{lyap} (denoted by $Z_p(t,\tau))$ is $Z_p(t,\tau)=Z_f(t)+Z_s(\tau) - \pi_0Z(0)\pi_0^T$, provides the long-time approximation to $Z$, the covariance of the stationary distribution, and is the sum of the fast and slow contributions
\begin{equation}\label{total}
\lim_{t,
\tau\to\infty}Z_p(t,
\tau) =: Z_p^{\infty} = \cfrac{B_0B_0^T}{2|\lambda_-^{(0)}|} + \cfrac{\pi_0B_1B_1^T\pi_0^T}{2|\lambda_+^{(1)}|}.
\end{equation}
Again, the first term on the right hand side of \eqref{total} accounts for the diffusion on the fast timescale $t$, while the second term accounts for diffusion that occurs on the slow timescale, $\tau$.

With the formulation of the composite expansion, we are now in a position to comment on the accuracy of the projected LNA as it pertains to the covariance.
\begin{proposition}\label{prop3}
Suppose $Z(0)=0$. The covariance of the projected LNA \eqref{SDE0P} converges to the covariance of the complete LNA \eqref{SDE0} as $\varepsilon \to 0$ if $\pi_0B_0=0$ and diffusion occurs only on the slow timescale. 
\end{proposition}
\begin{proof}
With $Z(0)=0$, the covariance  of the projected LNA \eqref{SDE0P}, given by $Z_p$, is 
\begin{equation}\label{PC}
Z_p(\tau) = \cfrac{\pi_0B_1B_1^T\pi_0^T}{2|\lambda_+^{(1)}|}(1-\exp(2\lambda_+^{(1)}\tau)),
\end{equation}
which is exactly the outer solution, $Z_s$, with $Z(0)=0$. From Fenichel's theorem, $Z(t) \to Z_s(t)$ as $\varepsilon \to 0$ if $Z(0)=0$ and $B_0 =0.$ 

Now suppose $B_0 \neq 0$. The steady-state covariance, $\lim_{t\to \infty} Z(t) := Z^{\infty}$, converges to $Z^{\infty}_0$ as $\varepsilon \to 0$:\footnote{The proof of this statement can be found in the Appendix. See Proposition \ref{prop4}.}
\begin{equation}
\lim_{\varepsilon \to 0} Z^{\infty} = Z^{\infty}_0= \cfrac{B_0B_0^T}{2|\lambda_-^{(0)}|} + \cfrac{\pi_0B_1B_1^T\pi_0^T}{2|\lambda_+^{(1)}|}.
\end{equation}
The covariance of the projected LNA is still given by \eqref{PC}. However, the steady-state covariance of $Z_p(\tau)$ is
\begin{equation}
\lim_{\tau \to \infty}Z_p(\tau) = \cfrac{\pi_0B_1B_1^T\pi_0^T}{2|\lambda_+^{(1)}|}=:Z_p^{\infty}
\end{equation}
and thus $Z_p^{\infty}\neq Z^{\infty}_0$. 
\end{proof}

There are several takeaways from this analysis. First, in contrast to the mean, we cannot mitigate discrepancies between $Z(t)$ and $Z_p(t)$ by choosing an appropriate initial condition (i.e., analogous to choosing $\mathbb{E}\{Y(0)\}=0$), or by allowing $t\to \infty$, since the projected LNA \eqref{SDE0P} does not account for any diffusion that occurs on the fast timescale. Thus, the covariance obtained from the projected LNA, \eqref{SDE0P}, will provide a reliable approximation to the covariance of the full LNA \eqref{SDE0} whenever diffusion occurs entirely on the slow timescale. 

Second, when diffusion is limited to the slow timescale and $B_0$ is identically zero, the LNA assumes the form
\begin{equation}\label{SDED}
{\rm d}Y = (A_0 + \varepsilon A_1) Y \;{\rm d}t + \sqrt{\varepsilon}\cdot B(x^*)\; {\rm d}W(t).
\end{equation}
By employing the Brownian motion scaling law\footnote{The notation ``$\myeq$" denotes distributional equality. }
\begin{equation}\label{ScalingLaw}
\cfrac{1}{\sqrt{\varepsilon }} W(\varepsilon t) \myeq W(t),
\end{equation}
\eqref{SDED} becomes
\begin{equation}\label{SDED1}
{\rm d}Y = A_0Y {\rm d}t +  A_1 Y \;{\rm d}\tau + \ B(x^*)\; {\rm d}W(\tau),
\end{equation}
and therefore the evolution of $Y$ on the fast timescale, $t$, is -- to leading order in $\varepsilon$ -- completely deterministic,
\begin{equation*}
\dot{Y}= A_0Y,
\end{equation*}
which implies that $\mathbb{E}\{Y\}$ will provide a reasonably good approximation to the LNA along the approach to the slow eigenspace, provided $\varepsilon$ is sufficiently small.

Third, the Fredholm alternative \eqref{FAL} does not always hold; this can happen, for example, when the underlying mass action system is perturbed {\it regularly} rather than singularly. We illustrate the implications of the Fredholm alternative, as well as the behavior of the LNA over fast timescales, with two examples presented in Subsection \ref{Rm}.

\subsection{Didactic examples}\label{Rm}

In this subsection, we present two examples involving the linear reaction network
\begin{align}\label{LRN}
    \emptyset \ce{->[$k_0$] U},\qquad\ce{U <=>[$k_1$][$k_{2}$] W ->[$k_3$] Z}.
\end{align}
Let lowercase $u,w,$ and $z$ denotes the respective concentration of $U,W$ and $Z$. The mass action equations that describe the temporal evolution of $u,w$ and $z$ are:
\begin{subequations}\label{linMASS}
\begin{align}
\dot{u} &= k_0 -k_1 u + k_2w,\\
\dot{w} &= k_1u - (k_2+k_3)w.
\end{align}
\end{subequations}
The system \eqref{linN} admits the stationary point, $(u^*,w^*)$:
\begin{equation}
u^* = \cfrac{k_0(k_2+k_3)}{k_1k_3}, \quad w^* = \cfrac{k_0}{k_3}
\end{equation}
In the examples the follow, we will consider two different perturbed versions of \eqref{linMASS}. In the first example, we will analyze a singularly perturbed form of \eqref{linMASS} whose corresponding LNA diffuses entirely on the slow timescale. In the second example, we will analyze a regularly perturbed version of \eqref{linMASS} for which the Fredholm alternative \eqref{FAL} fails to hold. 

\begin{example}
In this example we treat $k_0$ and $k_1$ as small parameters:\footnote{For a thorough definition of singular perturbation parameters for CRNs see~\citet{GOEKE2015}. }
\begin{equation}\label{map1}
(k_0,k_1) \mapsto (\varepsilon k_0,\varepsilon k_1),
\end{equation}
and analyze the limiting behavior of the LNA as $\varepsilon \to 0$. The modified mass action equations under \eqref{map1} are
\begin{subequations}\label{linN}
\begin{align}
\dot{u} &= \varepsilon k_0 -\varepsilon k_1 u + k_2w,\\
\dot{w} &= \varepsilon k_1u - (k_2+k_3)w,
\end{align}
\end{subequations}
and the stationary point $(u^*,w^*)$ assumes the form
\begin{equation}
u^* = \cfrac{k_0(k_2+k_3)}{k_1k_3}, \quad w^* = \cfrac{\varepsilon k_0}{k_3} =: \varepsilon \bar{w},\quad \bar{w}=k_0/k_3.
\end{equation}
Note that $w^*$ is $\mathcal{O}(\varepsilon)$.

Applying the Brownian motion scaling law and expressing the LNA in terms of $\tau$ and $t$ yields
\begin{subequations}\label{LNA0}
\begin{align}
{\rm d}y_1 &= -k_1 y_1 \;{\rm d}\tau+ k_2y_2 \;{\rm d}t + \sqrt{k_0}\;{\rm d}W_1(\tau) - \sqrt{k_1 u^*}\;{\rm d}W_2(\tau) + \sqrt{k_2 \bar{w}}\;{\rm d}W_3(\tau),\\
{\rm d}y_2 &= k_1 y_1 \;{\rm d}\tau- (k_2+k_3)y_2 \;{\rm d}t+ \sqrt{k_1 u^*}\;{\rm d}W_2(\tau)  -\sqrt{k_2 \bar{w}}\;{\rm d}W_3(\tau) - \sqrt{k_3 \bar{w}}\;{\rm d}W_4(\tau),
\end{align}
\end{subequations}
from which it is very clear that all diffusion occurs on the slow timescale, $\tau=\varepsilon t$, with drift occurring on both $t$ and $\tau$. The center subspace, $E^0$, is simply the $y_1$-coordinate axis: $E^0:=\{(y_1,y_2)\in \mathbb{R}^2: y_2=0\}$. If we start sufficiently far away from $E^0$ then, over fast timescales, realizations of the LNA are well-approximated by the expectation, $\mathbb{E}\{Y\}$. Moreover, we can approximate $\mathbb{E}\{y_1\}$ from the composite expansion,
\begin{equation}\label{compY1}
y_1(t,\tau) \approx -\cfrac{k_2}{k_2+k_3}y_2(0)\exp(-(k_2+k_3)t) +\left[y_1(0) + \cfrac{k_2}{k_2+k_3}y_2(0)\right]\exp(- \cfrac{k_1k_3}{k_2+k_3}\tau).
\end{equation}
See {\sc figure} \ref{FIG2} for a numerical illustration.
\begin{figure}[htb!]
\centering
\includegraphics[width=10.0cm]{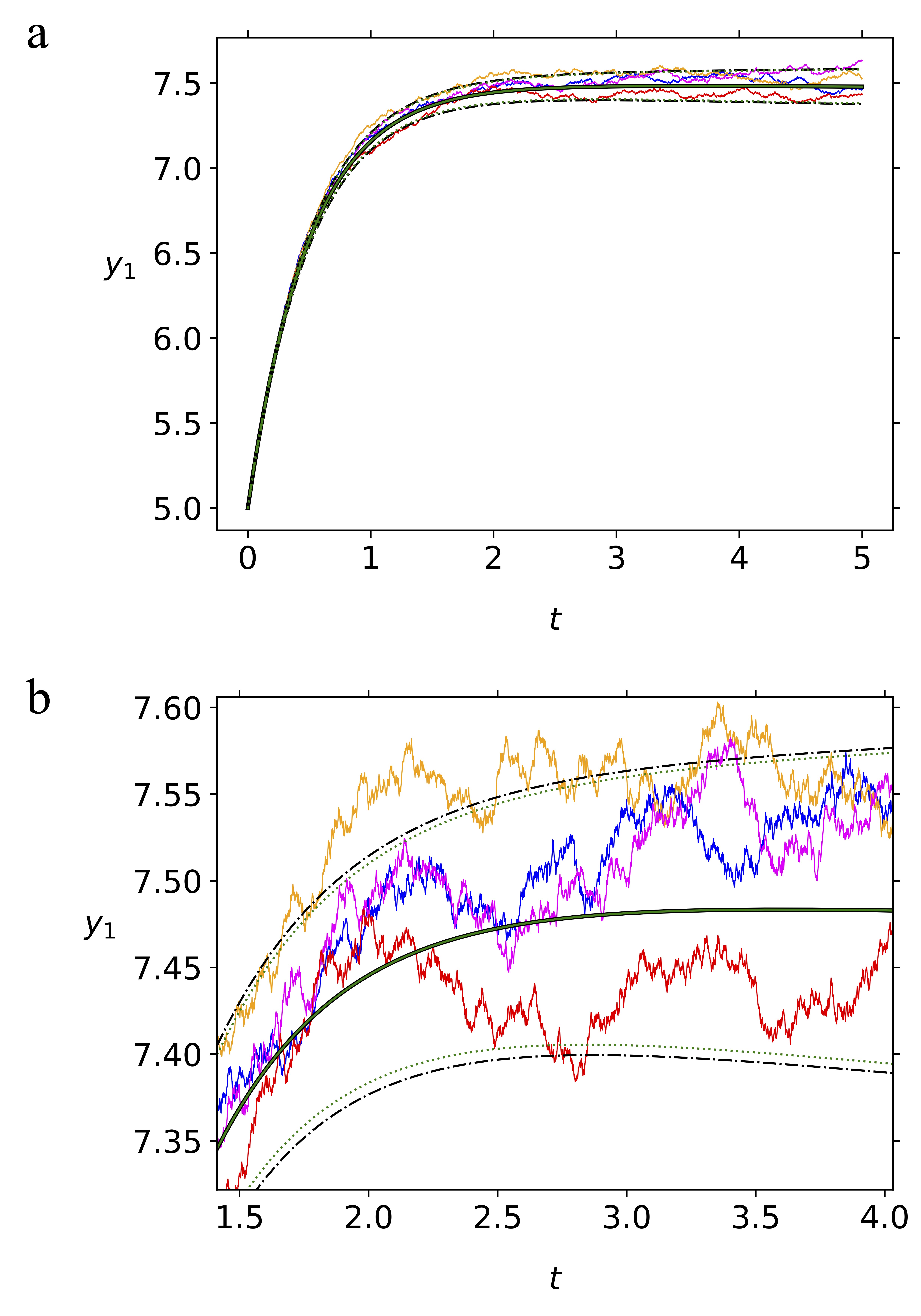}
\caption{\textbf{When diffusion occurs on the slow timescale, the approach to the slow eigenspace can be approximated with the expectation, $\mathbb{E}\{Y\}$, which obeys a deterministic ordinary differential equation.} In both panels the thick, solid curve is the expected value, $\mathbb{E}\{Y\}$, and the dashed/dotted curves are the mean $\pm$ the standard deviation obtained from the full LNA \eqref{LNA0}. The thick green line is the approximate mean for $y_1$, obtained from the composite solution \eqref{compY1}, and the dotted green curves are the composite solution for $y_1$, $\pm$ the standard deviation obtained from projected LNA \eqref{redcov}. The red, blue, orange and magenta curves are numerically-integrated realizations of the LNA \eqref{LNA0}. Parameter values used in the simulations (obtained via numerical integration of \eqref{LNA0}) are (in arbitrary units): $k_0=k_1=k_2=k_3=1.0$, $\varepsilon = 0.001$, and $y_1(0)=y_2(0)=5.0$. Panel a: Note that the LNA realizations do not deviate significantly from $\mathbb{E}\{Y\}$ during transient decay; only after the decay of transients do we start to see significant deviations from the expectation due to influence of diffusion. Moreover, the variance obtained from the projected LNA is highly accurate. Panel b: A close-up of the panel a.} \label{FIG2}
\end{figure}

Observe from \eqref{LNA0} that $B_0$ is identically zero, since all diffusion occurs on the slow timescale. Moreover, we also see from \eqref{LNA0} that
\begin{equation*}
A_0 = \begin{pmatrix}0 & k_2 \\ 0 & -(k_2+k_3)\end{pmatrix}, \quad \ker A_0 = {\rm  span} \bigg\{\begin{pmatrix}1\\ 0\end{pmatrix}\bigg\} =: E^0, \qquad A_1 = k_1\begin{pmatrix}-1&0\\\;\;\;1&0\end{pmatrix}.
\end{equation*}
Since $B_0=0$, the critical manifold $\mathcal{M}_0$ is identical to the center subspace of $\mathcal{L}_0$:
\begin{equation}
\ker \mathcal{L}_0(Z) = {\rm span} \bigg\{\begin{pmatrix}1&0\\0&0\end{pmatrix}\bigg\} =: \mathcal{M}_0, \quad B_1 = \begin{pmatrix}\sqrt{k_0} & -\sqrt{k_1u^*}& \sqrt{k_2\bar{w}} & 0\\ 0 & \sqrt{k_1u^*}&  -\sqrt{k_2\bar{w}} &  -\sqrt{k_3\bar{w}}\end{pmatrix},
\end{equation}
and the covariance equation satisfies
\begin{equation}\label{PJ1}
\dot{Z} = \mathcal{L}_0(Z) + \varepsilon \mathcal{L}_1(Z) + \varepsilon B_1B_1^T.
\end{equation}
The projected form of \eqref{LNA0} is 
\begin{multline}\label{LNA0P}
{\rm d}y_{1p} = -\cfrac{k_1k_3}{k_2+k_3}\cdot y_{1p}\; {\rm d}\tau+ \sqrt{k_0}\;{\rm d}W_1(\tau) - \cfrac{k_3}{k_2+k_3}\cdot \sqrt{k_1 u^*}\;{\rm d}W_2(\tau)\\ + \cfrac{k_3}{k_2+k_3}\sqrt{k_2 \bar{w}}\;{\rm d}W_3(\tau)- \cfrac{k_2}{k_2+k_3}\sqrt{k_3\bar{w}}\;{\rm d}W_4(\tau),\\
\end{multline}
with ${\rm d}y_{2p}=0$. The variance of $y_{1p}$, $\widehat{z}_{11}$, obtained the projected form of \eqref{PJ1},
\begin{equation*}
\dot{Z}_p= \varepsilon \pi_0 \mathcal{L}_1(Z_p)\pi^T + \varepsilon \pi_0B_1B_1^T\pi_0^T,\quad Z_p=\begin{pmatrix}\widehat{z}_{11}& \widehat{z}_{12} \\ \widehat{z}_{21} & \widehat{z}_{22} \end{pmatrix}
\end{equation*}
is given by the ordinary differential equation
\begin{equation}\label{redcov}
\cfrac{{\rm d}\widehat{z}_{11}}{{\rm d} \tau} = -\cfrac{2k_1k_3}{k_2+k_3} \widehat{z}_{11} + k_0 + \left(\cfrac{k_3}{k_2+k_3}\right)^2\left(k_1u^*+k_2\bar{w}\right) + \left(\cfrac{k_2}{k_2+k_3}\right)^2\cdot k_3\bar{w},
\end{equation}
with the additional covariance equations given by
\begin{equation*}
\cfrac{{\rm d} \widehat{z}_{21}}{\rm{d}\tau} = \cfrac{{\rm d} \widehat{z}_{12}}{\rm{d}\tau}=\cfrac{{\rm d} \widehat{z}_{22}}{\rm{d}\tau} = 0.
\end{equation*}
Furthermore, $\widehat{z}_{12}=\widehat{z}_{12} = \widehat{z}_{22} =0$ on $\mathcal{M}_0$, and the only variable whose flow is nontrivial on $\mathcal{M}_0$ is the variance of $y_{1p}$, $\widehat{z}_{11}$. It follows from Fenichel theory that $z_{11}$ is well-approximated by $\widehat{z}_{11}$. Moreover, the long-time covariance of the full LNA, $Z^{\infty}$, converges to the long-time covariance of the projected LNA, $Z_p^{\infty}$,
\begin{equation}
Z_p^{\infty} = \begin{pmatrix}\widehat{z}_{11}^{\infty}&0\\0&0\end{pmatrix}, \quad \widehat{z}_{11}^{\infty}=: \cfrac{k_0 + \left(\cfrac{k_3}{k_2+k_3}\right)^2(k_1u^* + k_2\bar{w}) +\left(\cfrac{k_2}{k_2+k_3}\right)^2k_3\bar{w}}{\cfrac{2k_1k_3}{k_2+k_3}}
\end{equation}
as $\varepsilon \to 0$; again, see {{\sc figure}} \ref{FIG2} for a numerical illustration.
\end{example}
In our next example, we illustrate what can happen when the Fredholm alternative \eqref{FAL} fails to hold and $\pi_0B_0\neq 0$. 
\begin{example}
To understand what can happen when \eqref{FAL} fails to hold, consider once again the reaction network \eqref{LRN} but with a $k_0$ that is $\mathcal{O}(1)$:
\begin{equation*}
(k_0,k_1) \mapsto (k_0,\varepsilon k_1).
\end{equation*}
In this case, the mass action equations are
\begin{subequations}\label{linN2}
\begin{align}
\dot{u} &= k_0 -\varepsilon k_1 u + k_2w,\\
\dot{w} &= \varepsilon k_1u - (k_2+k_3)w,
\end{align}
\end{subequations}
which is a {\bf regularly perturbed}\footnote{In this case, setting $\varepsilon =0$ results in the invariance of the $u$-axis ($w=0$), but the resulting $\mathcal{O}(1)$ system is void of stationary points and is therefore not a singular perturbation; see~\citet{OpenMMin} for details.} differential equation system. The LNA about the stationary point is 
\begin{subequations}\label{LNA2}
\begin{align}
{\rm d}y_1 &= -k_1 y_1 \;{\rm d}\tau+ k_2y_2 \;{\rm d}t + \sqrt{k_0}\;{\rm d}W_1(t) - \sqrt{k_1 u^*}\;{\rm d}W_2(t) + \sqrt{k_2 \bar{w}}\;{\rm d}W_3(t),\\
{\rm d}y_2 &= k_1 y_1 \;{\rm d}\tau- (k_2+k_3)y_2 \;{\rm d}t+ \sqrt{k_1 u^*}\;{\rm d}W_2(t)  -\sqrt{k_2 \bar{w}}\;{\rm d}W_3(t) - \sqrt{k_3 \bar{w}}\;{\rm d}W_4(t).
\end{align}
\end{subequations}
Observe that while the drift terms in \eqref{LNA0} and \eqref{LNA2} are identical, the latter contains diffusive terms that evolve on the fast timescale since $k_0, k_1u^*$ and $\bar{w}$ are all $\mathcal{O}(1)$. There are several consequences. First, realizations of \eqref{LNA2} can immediately -- and significantly -- depart from $\mathbb{E}\{Y\}$ since diffusion occurs on the fast timescale; see {{\sc figure}} \ref{FIG3} for a numerical illustration.
\begin{figure}[htb!]
\centering
\includegraphics[width=10.0cm]{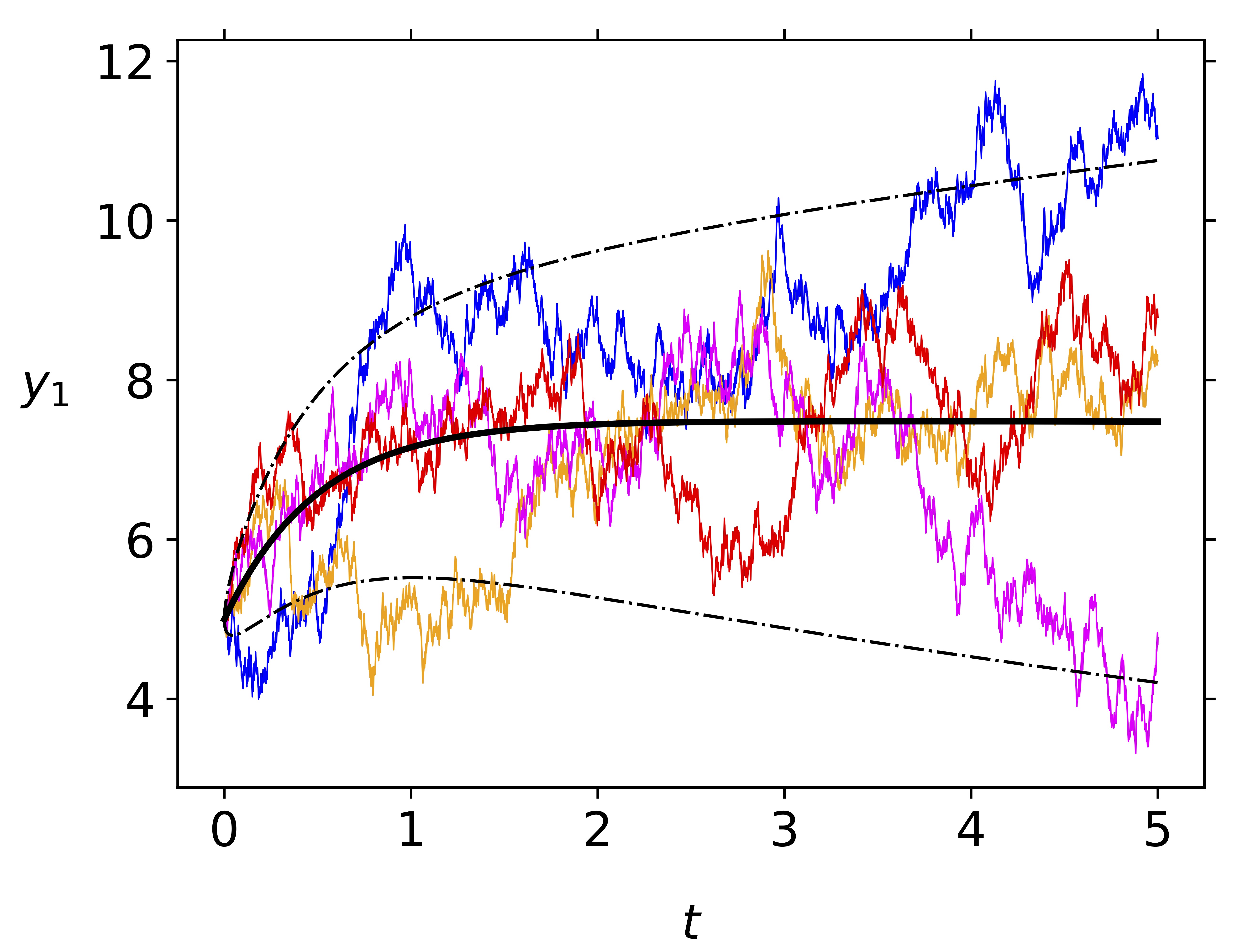}
\caption{\textbf{When diffusion occurs on the fast timescale, realizations of the LNA can depart from the mean immediately.} The thick, solid back curve is the expected value, $\mathbb{E}\{Y\}$, and the thin, dashed/dotted black curves are the mean $\pm$ the standard deviation. The red, blue, orange, and magenta curves are numerically integrated realizations of the LNA \eqref{LNA0}. The parameter values used in the simulations (obtained through the numerical integration of \eqref{LNA2}) are (in arbitrary units): $k_0=k_1=k_2=k_3=1.0$, $\varepsilon = 0.001$, and $y_1(0)=y_2(0)=5.0$. Note that the LNA realizations deviate significantly from $\mathbb{E}\{Y\}$ during transient decay due to diffusion that occurs on the fast timescale. This behavior is markedly different than the behavior presented in {{\sc figure}} \ref{FIG2}.} \label{FIG3}
\end{figure}

Second, the matrices $B_0$ and $B_1$ are given by
\begin{equation*}
B_0 =\begin{pmatrix}\sqrt{k_0} & -\sqrt{k_1u^*}&\;\;\;\sqrt{k_2\bar{w}} & 0\\ 0 & \sqrt{k_1u^*} & -\sqrt{k_2\bar{w}} & -\sqrt{k_3\bar{w}}\end{pmatrix}, \quad B_1=0
\end{equation*}
It is straightforward to verify that \eqref{FAL} fails to hold, and this adversely impacts the steady-state covariance, $Z^{\infty}$, in the limit as $\varepsilon \to 0$: Recall that the Lyapunov operator $\mathcal{L}(Z) = A(\varepsilon)Z+ZA^T(\varepsilon)$ is invertible when $\varepsilon \neq 0$. The steady-state covariance, $Z^{\infty}$, is given by
\begin{equation*}
Z^{\infty}=-\mathcal{L}^{-1}(BB^T).
\end{equation*}
However, 
\begin{equation}
\lim_{\varepsilon \to 0}||Z^{\infty}||\quad \text{{\rm does not exist}}
\end{equation}
due to the presence of $\mathcal{O}(1)$ components of $BB^T$ that lie within $\ker \mathcal{L}(\cdot)$ as $\varepsilon \to 0$; see Appendix, Proposition \ref{prop5}, for a formal proof of this statement. The consequence is that instead of settling down to a limiting steady-state covariance as $\varepsilon \to 0$, the variance of $y_1$ increases without bound as $\varepsilon \to 0$, and therefore realizations of \eqref{LNA0} tend to depart substantially from the expectation as $t\to \infty$ and $\varepsilon \to 0$.% 
\end{example}

Finally, it should be noted that while the covariance obtained from the projected LNA \eqref{SDE0P} will not approximate the covariance of the LNA unless $B_0=0$, it can still accurately approximate the covariance of individual components. For example, consider the {\it standard form}
\begin{subequations}\label{stochSF}
\begin{align}
{\rm d}y_1 &= \varepsilon \sum_{j=1}^2 A_{1j}Y_j\;{\rm d}t + \sqrt{\varepsilon}\cdot\sum_{m =1}^k B(x^*)_{1m}\;{\rm d}W_m(t),\label{rSFa}\\
{\rm d}y_2 &= \;\;\;\sum_{j=1}^2 A_{2j}Y_j\;{\rm d}t + \quad\phantom{\gamma}\sum_{m =1}^k B(x^*)_{2m}\;{\rm d}W_m(t),
\end{align}
\end{subequations}
where from inspection we see
\begin{equation}
A = A_0 + \varepsilon A_1 = \begin{pmatrix} 0 & 0\\ a_{21}& a_{22}
\end{pmatrix} + \varepsilon \begin{pmatrix} a_{11} & a_{12} \\ 0 & 0
\end{pmatrix}.
\end{equation}
Rewriting the first component \eqref{rSFa} in terms of the slow time, $\tau$, yields
\begin{equation}
{\rm d}y_1 = \sum_{j=1}^n A_{1j}Y_j\;{\rm d}\tau + \sum_{m =1}^k B(x^*)_{1m}\;{\rm d}W_m(\tau),
\end{equation}
and it is obvious that not only is $y_1$ effectively deterministic on the fast timescale $t$, it is also approximately constant. The reduction of \eqref{stochSF} involves the substitution, $y_2 = -a_{21}y_1/a_{22}$, into \eqref{rSFa}; this yields
\begin{equation}\label{stochred}
{\rm d}y_{1p} = \left(a_{11}-a_{12}\cdot \cfrac{a_{21}}{a_{22}}\right)y_{1p}\;{\rm d}\tau + \sum_{m=1}^k B(x^*)_{1m}\;{\rm d}W_m(\tau).
\end{equation}
Since both the drift and the diffusion of $y_1$ unfold on the slow timescale, the variance of $y_{1p}$, denoted by $z_{1p}$, obtained from \eqref{stochred}
\begin{equation}
z_{1p}' = 2\left(a_{11}+a_{12}\cdot \mu \right)z_{1p} + \sum_{m=1}^k \left[B(x^*)_{1m}\right]^2, \quad \mu =: -a_{21}/a_{22}
\end{equation}
is a very good approximation to $z_1$, the variance of $y_1$ obtained from the full (unprojected) LNA. We will not prove this statement here since this result -- which pertains to systems in the standard form \eqref{stochSF} -- is rather well-established in the literature; see~\cite{Yin2004,Yin2005,PAHLAJANI201196,Herath,BerglundGentzJDE,JSEssLNA}.

On the other hand, the projected form of $y_2$ is 
\begin{equation}
{\rm d}y_{2p} = \mu \left(a_{11}+a_{12}\mu \right)y_{1p}\;{\rm d}\tau +\mu \sum_{m=1}^k B(x^*)_{1m}\;{\rm d}W_m(\tau) = \mu {\rm d}y_{1p},\label{redB}
\end{equation}
but \eqref{redB} only approximates the drift and diffusion of $y_2$ on the slow timescale. Consequently, the variance of $y_{2p}$ determined by \eqref{redB} will not approximate $z_2$ since \eqref{redB} neglects the diffusion of $y_2$ on the fast timescale.

The takeaway is that, while it may not always be possible to estimate the long-time covariance of the LNA by projection onto $E^0$, we can approximate the first and second moments of the {\it components} of the LNA whose drift and diffusion are negligible over fast timescales. In Section 5 we look at several examples from the literature that illustrate these concepts.

\section{Model reduction strategies, revisited.}

In this section, we take a close look at two reduction strategies from the literature. The first is the slow-scale linear noise approximation formulated by~\citet{Thomas2012}, which has been shown to provide a highly accurate reduced LNA when applied to enzymatic reaction networks. The second is the {\it total quasi-steady-state approximation} which, based on the results of numerical simulations, has been reported to be an effective reduction technique for enzymatic reactions; see~\cite{KIM2014,KIM2015,Kim2020,Burrage2008}. In both examples we consider the {\it open} Michaelis-Menten reaction,
\begin{align}\label{mm1}
    \emptyset \ce{->[$k_0$] S},\qquad\ce{S + E <=>[$k_1$][$k_{-1}$] C ->[$k_2$] E + P},
\end{align}
where $k_0$, $k_1$, $k_{-1}$ and $k_2$ are rate constants. Let $s$ and $c$ denote the concentrations of substrate, $S$, and complex, $C$, respectively. The mass action ODE system that describes the deterministic evolution of concentrations is
\begin{equation}
\begin{pmatrix}
\dot{s}\\\dot{c}
\end{pmatrix} = \begin{pmatrix}1&-1&\;\;\;1&\;\;\;0\\0&\;\;\;1&-1&-1\end{pmatrix}\begin{pmatrix}k_0\\k_1(e_0-c)s\\k_{-1}c\\k_2c\end{pmatrix}=: Sr(s,c),
\end{equation}
where $e_0$ is the total enzyme concentration (free and bound) and is a conserved quantity. We will take the influx rate of substrate, $k_0$ to be $\alpha k_2e_0$, where $\alpha\in [0,1)$ and ensures that the system has a non-trivial fixed point, $(s^*,c^*)$ in the first quadrant located at 
\begin{equation}
c^* = \alpha e_0, \quad s^* = \alpha K_M/(1-\alpha), 
\end{equation}
where $K_M=(k_{-1}+k_2)/k_1$ is the Michaelis constant. 

\subsection{The total quasi-steady-state approximation}
The total quasi-steady approximation (tQSSA) is a reduction method commonly employed to reduce enzyme reaction networks. Although the mechanism behind the success of the {\it deterministic} tQSSA is understood~\cite{Unreasonable}, the accuracy of its stochastic counterpart is an area of active research; see~\citet{ganguly2025asymptotic,Kang2019}. We turn to enzyme kinetics to understand {\it why} the tQSSA is effective. 

\begin{example}
Consider small $k_2$, and set $k_2\mapsto \varepsilon k_2$ and $k_0\mapsto \varepsilon k_0$ since $k_0=\alpha k_2e_0$. The LNA in this case is
\begin{multline}\label{LNAk2}
\begin{pmatrix}
{\rm d}y_s\\{\rm d}y_c
\end{pmatrix} = \begin{pmatrix}-k_1e_0(1-\alpha) & \;(\alpha \varepsilon k_2 + k_{-1})/(1-\alpha)\\ \;\;\;k_1e_0(1-\alpha) & -(k_{-1}+\varepsilon k_2)/(1-\alpha)\end{pmatrix}\begin{pmatrix}y_s\\y_c\end{pmatrix}{\rm d}t + \\
\begin{pmatrix}
\sqrt{\alpha \varepsilon k_2 e_0}&-\sqrt{\alpha e_0(k_{-1}+\varepsilon k_2)} & \sqrt{\alpha k_{-1}e_0} & 0\\
 0& \sqrt{\alpha e_0(k_{-1}+\varepsilon k_2)} & - \sqrt{\alpha k_{-1}e_0} & -\sqrt{\alpha k_{2}\varepsilon e_0}
\end{pmatrix}\begin{pmatrix}{\rm d}W_1(t)\\{\rm d}W_2(t)\\{\rm d}W_3(t)\\{\rm d}W_4(t)\end{pmatrix}.
\end{multline}
Setting $\varepsilon =0$ yields
\begin{multline}\label{F2}
\begin{pmatrix}
{\rm d}y_s\\{\rm d}y_c
\end{pmatrix} = \begin{pmatrix}-k_1e_0(1-\alpha) &  \;\;k_{-1}/(1-\alpha)\\ \;\;\;k_1e_0(1-\alpha) & -k_{-1}/(1-\alpha)\end{pmatrix}\begin{pmatrix}y_s\\y_c\end{pmatrix}{\rm d}t + \\\sqrt{\alpha e_0k_{-1}}\cdot
\begin{pmatrix}-1 & \;\;\;1 \\
\;\;\;1 & - 1 
\end{pmatrix}\begin{pmatrix}{\rm d}W_2(t)\\{\rm d}W_3(t)\end{pmatrix},
\end{multline}
from which we see an immediate obstacle: the fast subsystem \eqref{F2} has both drift and diffusion terms that evolve on the fast timescale. The long-time variance of $Y$, $Z^{\infty}$, converges to 
\begin{equation}
\lim_{\varepsilon \to 0} Z^{\infty} = Z^{\infty}_0 = \cfrac{B_0B_0^T}{2|\lambda_-^{(0)}|} + \cfrac{\pi_0B_1B_1^T\pi_0^T}{2|\lambda_+^{(1)}|} 
\end{equation}
as $\varepsilon \to 0 $, where the leading order approximations to the eigenvalues of $A$ are
\begin{equation}
\lambda_-^{(0)}=-\cfrac{k_1e_0(1-\alpha)^2+k_{-1}}{(1-\alpha)}, \quad \lambda_+^{(1)} = -\cfrac{k_2k_1e_0(1-\alpha)^2}{k_1e_0(1-\alpha)^2+k_{-1}}
\end{equation}
and the matrices $B_0$ and $B_1$ are given by\footnote{The matrix $B_0$ is constructed with the $\mathcal{O}(1)$ approximation to the stationary point, $s^*=\alpha K_S/(1-\alpha)$, instead of $s^*=\alpha K_M/(1-\alpha)$.}
\begin{equation}
B_0 = \sqrt{\alpha k_{-1}e_0}\cdot\begin{pmatrix}-1& \;\;\; 1\\\;\;\;1&-1\end{pmatrix},\quad B_1 = \sqrt{\alpha\varepsilon k_2e_0}\cdot \begin{pmatrix}1& \;\;\;0\\ 0 & -1\end{pmatrix}.
\end{equation}

In its entirety, the long-time variance with $\varepsilon =0$, $Z^{\infty}_0$, is
\begin{equation}\label{zinfty}
 Z^{\infty}_0=\begin{pmatrix}\cfrac{k_{-1}\alpha(k_1e_0(1-\alpha)^3+k_{-1})}{k_1(k_1e_0(1-\alpha)^2+k_{-1})(1-\alpha)^2} & \cfrac{k_{-1}\alpha^2e_0}{{k_1e_0(1-\alpha)^2+k_{-1}}} \\ \cfrac{k_{-1}\alpha^2e_0}{{k_1e_0(1-\alpha)^2+k_{-1}}} & \cfrac{(1-\alpha)\alpha e_0 (k_1e_0(1-\alpha)+k_{-1})}{{k_1e_0(1-\alpha)^2+k_{-1}}}\end{pmatrix}
\end{equation}
and accounts for diffusion on the fast and slow timescales. However, if we were to simply project \eqref{LNAk2} onto $E^0$, the center subspace of $A_0$, we would estimate the long-time variance to be 
\begin{equation*}
\cfrac{\pi_0B_1B_1^T\pi_0^T}{2|\lambda_+^{(1)}|},
\end{equation*}
which is inaccurate since it underestimates \eqref{zinfty} by a difference of $B_0B_0^T/2|\lambda_-^{(0)}|$; see {{\sc figure}} \ref{FIG4} for a numerical illustration.

Thus, although we can reduce the LNA \eqref{LNAk2} simply by projecting onto $E^0$, this projection will eliminate the influence of diffusion on the fast timescale and underestimate the long-term variance. Again, this is because $y_s$ and $y_c$ evolve on both the fast and slow timescales. However, this raises an important question: Can we at least find a coordinate transformation that transforms \eqref{LNAk2} into the standard form \eqref{stochSF}?

Looking carefully at the fast subsystem \eqref{F2}, we see that summing the components together yields
\begin{equation}
{\rm d}y_s + {\rm d}y_c = {\rm d }(y_s+y_c) = {\rm d}y_T=0,
\end{equation}
where the sum, $y_s+y_c = y_T$, is called the  ``total" substrate. Thus, the total substrate, $y_T$, is effectively constant on the fast timescale, which means that the projected form 
\begin{equation}\label{totals}
{\rm d}y_{Tp} = \lambda_+^{(1)}y_{Tp}\;{\rm d}\tau + \sqrt{k_0}\;{\rm d}W_1(\tau)-\sqrt{\alpha k_2e_0}\;{\rm d}W_4(\tau)
\end{equation}
will adequately approximate the expectation and variance of the total substrate since both the drift and diffusion unfold on the slow timescale. Moreover, it is straightforward to check that the long-time variance of the total substrate obtained from \eqref{totals} is
\begin{equation}
\mathbb{V}\{y_{Tp}\} = \cfrac{\alpha(k_1e_0(1-\alpha)^2+k_{-1})}{k_1(1-\alpha)^2},
\end{equation}
which is exactly the sum of the entries of $Z_0^{\infty}$ given by \eqref{zinfty}. Thus, $Z_0^{\infty}(y_T)\to Z_p^{\infty}(y_{Tp})$ as $\varepsilon \to 0$.
\end{example}

The takeaway from this example is twofold. First, it is sometimes possible to find a tractable coordinate transformation that brings the LNA into standard form \eqref{stochSF} with distinct slow and fast processes. In this case, it is always possible to generate an accurate reduced LNA for the slow variable that evolves entirely on the slow timescale. Second, this is precisely {\it why} the tQSSA works: When product formation and substrate influx are slow, both substrate and complex concentration are fast variables, which means that they can change significantly over fast {\it and} slow timescales. However, the addition of the complex and the substrate generates a new variable -- the {\it total} substrate -- which is entirely {\it slow} and is effectively constant over the fast timescale. In other words, the transformation to ``total" substrate coincides with a transformation to the standard form \eqref{stochSF}, with the total substrate defining the slow variable.

\subsection{The slow-scale Linear Noise Approximation (ssLNA)}

The slow-scale linear noise approximation (ssLNA) was originally derived by~\citet{Thomas2012} and is highly accurate when applied to various gene expression and enzymatic networks. Specifically, for enzymatic networks, the ssLNA is known to accurately approximate the mean and variance of substrate concentration when the enzyme concentration is sufficiently small. However, two key ingredients are missing from the original derivation of the ssLNA: First, the ssLNA was derived under the a priori assumption that enzymatic reactions with small enzyme concentration automatically assume the standard form \eqref{stochSF}. However, the ssLNA is noticeably different from the reduced LNAs of~\citet{PAHLAJANI201196} and~\citet{Herath} which were derived under the same a priori assumption. Second, the ssLNA's accuracy has been confirmed through numerical simulations and analyses of enzymatic reaction networks, but its derivation is not rooted in singular perturbation theory~\cite{ThomasPO}. Moreover, there is no formal proof that the first and second moments of the ssLNA converge to the first and second moments of the LNA $\varepsilon \to 0$. Our aim here is not to challenge the legitimacy of the ssLNA, as this is well-established; instead, our aim is to put the ssLNA on solid mathematical footing by establishing {\it when} and {\it why} the ssLNA is accurate when applied to enzymatic reactions with low enzyme concentration. 

\begin{example}
To understand the effectiveness of the ssLNA as applied to enzymatic networks, it suffices to reduce the LNA of the \textit{open} Michaelis-Menten reaction mechanism with $e_0 \mapsto \varepsilon e_0$, which implies $k_0 \mapsto \varepsilon k_0$, since $k_0=\alpha k_2e_0$. The LNA in this case is
\begin{multline}\label{LNAe0}
\begin{pmatrix}
{\rm d}y_s\\{\rm d}y_c
\end{pmatrix} = \begin{pmatrix}-k_1\varepsilon e_0(1-\alpha) & \;(\alpha k_2 + k_{-1})/(1-\alpha)\\ \;\;\;k_1\varepsilon e_0(1-\alpha) & -(k_{-1}+k_2)/(1-\alpha)\end{pmatrix}\begin{pmatrix}y_s\\y_c\end{pmatrix}{\rm d}t + \\
\begin{pmatrix}
\sqrt{\alpha k_2 \varepsilon e_0}&-\sqrt{\alpha \varepsilon e_0(k_{-1}+k_2)} & \sqrt{\alpha k_{-1}\varepsilon e_0} & 0\\
 0& \sqrt{\alpha \varepsilon e_0(k_{-1}+k_2)} & - \sqrt{\alpha k_{-1}\varepsilon e_0} & -\sqrt{\alpha k_{2}\varepsilon e_0}
\end{pmatrix}\begin{pmatrix}{\rm d}W_1(t)\\{\rm d}W_2(t)\\{\rm d}W_3(t)\\{\rm d}W_4(t)\end{pmatrix}.
\end{multline}

Note that with small enzyme concentration, all of the diffusion is limited to the slow timescale. Setting $\varepsilon =0$ yields the deterministic problem,
\begin{subequations}
\begin{align}
\dot{y}_s &= \cfrac{(\alpha k_2 +k_{-1})}{1-\alpha}y_c,\\
\dot{y}_c &= -\cfrac{(k_{-1}+k_2)}{1-\alpha}y_c,
\end{align}
\end{subequations}
from which it is clear that neither $y_s$ nor $y_c$ are inherently {\it slow}, as both can change over the fast timescale. However, since diffusion is limited to the slow timescale, we can reduce \eqref{LNAe0} via projection onto the center subspace, $y_c=0$. In its most general form, the projection onto the center subspace is
\begin{subequations}
\begin{align}
{\rm d}y_s &= -\cfrac{k_2e_0K_M}{(s^*+K_M)^2}\cdot y_s \;{\rm d}\tau + \sqrt{k_0}\;{\rm d}W_1(\tau) -\sqrt{\cfrac{k_2e_0s^*}{s^*+K_M}\left(1-\cfrac{2Ks^*}{(s^*+K_M)^2}\right)}{\rm d}W(\tau),\label{comp1}\\
{\rm d}y_c &= 0.\label{comp2}
\end{align}
\end{subequations}
where $K=:k_2/k_1$. The first component \eqref{comp1} is the ssLNA for the open MM reaction network \eqref{mm1}. The second component \eqref{comp2} is deterministic and therefore the variance of $y_c$ is identically zero. Moreover, from Proposition \ref{prop3}, the long-time covariance of the LNA, $Z^{\infty}$, converges to $Z_p^{\infty}$,
\begin{equation}
Z_p^{\infty}=\begin{pmatrix}\widehat{z}_{11}^{\infty} & 0\\ 0 & 0\end{pmatrix}, \qquad \widehat{z}_{11}^{\infty}= \cfrac{1}{2}\cdot \cfrac{k_0 + \cfrac{k_2e_0s^*}{s^*+K_M}\left(1-\cfrac{2Ks^*}{(s^*+K_M)^2}\right)}{\cfrac{k_2e_0K_M}{(s^*+K_M)^2}}
\end{equation}
as $\varepsilon \to 0$.
The key point from this example is that not only does the ssLNA approximate the variance in substrate concentration, it also approximates the variance of the complex concentration. This is because when the total enzyme concentration is small, the diffusion -- of both complex and substrate -- are limited to the slow timescale, and reduction of the entire LNA is achievable via projection onto the center subspace of the zeroth-order drift matrix, $A_0$. 
\end{example}

\section{Discussion}

We have shown that projecting LNA \eqref{LNA0} onto the center subspace (critical manifold) of the zeroth-order drift matrix, $A_0$, results in a reduced LNA \eqref{LNA0P} that accurately approximates the long-time expectation and covariance of the full LNA when diffusion is limited to the slow timescale. In this scenario, the evolution of the LNA over the fast timescale is, for all intents and purposes, deterministic. What then is the relationship between timescale separation and the accuracy of stochastic reductions in the linear noise regime? In most of the literature, timescale separation refers to a gap present in the eigenspectrum of the drift matrix. The expectation of the LNA, $\mathbb{E}\{Y\}$, satisfies a linear matrix equation,
\begin{equation*}
\dot{\mathbb{E}}\{Y\} = A\mathbb{E}\{Y\} ,
\end{equation*}
and the expectation eventually approaches the origin along the direction of the slow eigenvector. 

However, the behavior of the LNA is influenced not only by the eigenvalues of the drift matrix, but also by eigenvalues of the diffusion matrix. And, a gap in the drift matrix eigenspectrum does not necessarily imply the existence of a gap in the eigenspectrum of the diffusion matrix, $\mathcal{D} = \frac{1}{2}BB^T$. Because $\mathcal{D}$ is symmetric, it admits a pair of orthonormal eigenvectors, $u_1,u_2$ with 
\begin{equation}
u_i^Tu_j = \delta_{ij} \qquad \text{and}\quad \mathcal{D} = \mu_1u_1u_1^T + \mu_2u_2u_2^T,
\end{equation}
where $\mu_1,\mu_2$ are the eigenvalues of $\mathcal{D}$. If $\mu_1\sim \mathcal{O}(1)$ but $\mu_2\sim \mathcal{O}(\varepsilon)$, then a spectral gap is present and 
\begin{equation}
B_0B_0^T = 2\mu_1u_1u_1^T =:2\mathcal{D}_0,\qquad B_1B_1^T = \mu_2u_2u_2^T=:2\mathcal{D}_1(\varepsilon)\end{equation}
However, in some applications both $\mu_1$ and $\mu_2$ are $\mathcal{O}(\varepsilon)$, even though the corresponding drift matrix has one $\mathcal{O}(1)$ eigenvalue and one eigenvalue that is $\mathcal{O}(\varepsilon)$. This occurs, for example, in the case of the open Michaelis-Menten reaction with small enzyme concentration. The eigenvalues of diffusion matrix in this example are
\begin{equation}
\mu_1 = \varepsilon \alpha k_2e_0, \quad \mu_2 = 3\varepsilon \alpha k_2e_0 + 4\varepsilon \alpha k_{-1}e_0,
\end{equation}
which are both $\mathcal{O}(\varepsilon)$ and hence not disparate. Thus, if we equate timescale separation with {\it eigenvalue disparity}, then we have to specify what this implies since there are drift and diffusion timescales that must be considered. In fact, as we have shown, the lack of a gap in the spectrum of the diffusion matrix is exactly why the ssLNA of~\citet{Thomas2012} works so well for enzymatic reactions with low enzyme concentration.

\begin{figure}[htb!]
\centering
\includegraphics[width=10cm]{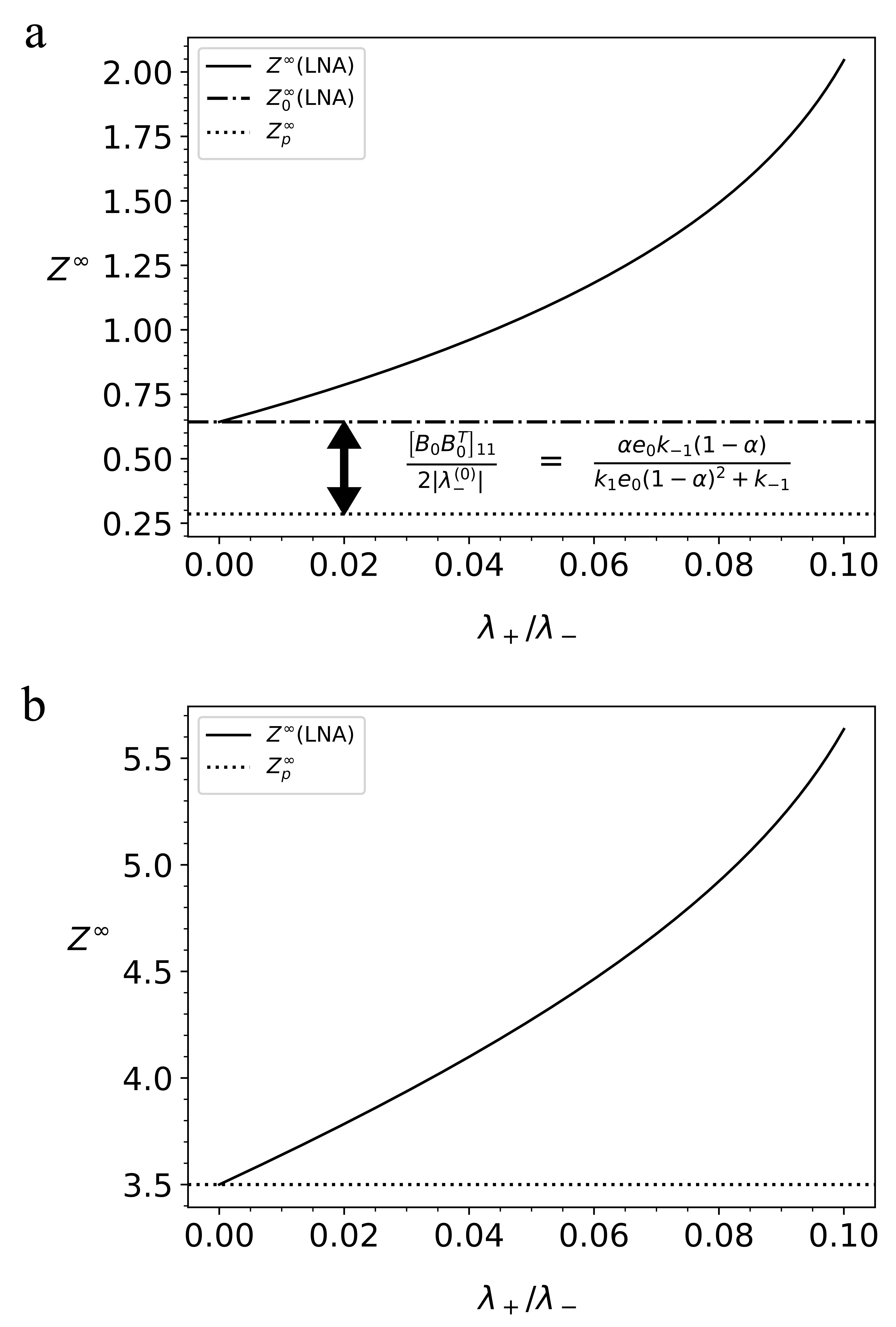}
\caption{\textbf{When diffusion occurs the fast and slow timescales, the projected LNA \eqref{SDE0P} will underestimate the variance of the LNA by a factor of $(1+\delta_0)$.} Panel a: The thick black curve is the limiting variance of substrate, $s$, as the eigenvalue ratio $\lambda_+/\lambda_-\to 0$ computed from the LNA of the open MM reaction mechanism discussed in {\bf Example 3.} The parameters used in each simulation are (in arbitrary units): $s_0=10.0,k_1=10.0,k_{-1}=5.0,\alpha=0.5$, and $e_0=5.0$. The catalytic rate constant, $k_2$, is varied from $0.0001$ to $10.0$ As the eigenvalue ratio vanishes, the limiting steady-state variance of substrate converges to $Z_0^{\infty}$ (dashed/dotted line). However, the projected LNA converges to $Z_p^{\infty}\neq Z_0^{\infty}$. The difference (bold arrows) is the variance of substrate resulting from fast timescale diffusion as $\varepsilon \to 0$, which is $\frac{\alpha e_0k_{-1}(1-\alpha)}{k_1e_0(1-\alpha)^2+k_{-1}}$. Panel b: The solid black curve is the steady-state covariance $Z^{\infty}$ of the {\it total substrate}, which converges to $Z_0^{\infty} = Z_p^{\infty}$ as the eigenvalue ratio vanishes. The coordinate transformation results in a new concentration (the total substrate) $s_T=s+c$, which diffuses on the slow timescale only. Consequently, the total substrate variance obtained from the projected LNA will converge to the total substrate variance obtained from the LNA as the eigenvalue ratio vanishes and $(\varepsilon,\delta)\to (0,0)$.}  \label{FIG4}
\end{figure}

More importantly, we are now able to answer to our original question regarding the accuracy of projecting the LNA onto the center manifold of the singular drift matrix, $A_0$. Let $\pi_{+}$ and $\pi_{-}$ denote the matrices that project onto the slow and fast eigenspaces of $A$. The variance of a particular component is 
\begin{multline}
z_{ii}^{\infty}(\varepsilon) =  \cfrac{\sum_j\left[\pi_+(\varepsilon)\mathcal{D}(\varepsilon)\pi_+^T(\varepsilon)\right]_{ij}}{|\lambda_+(\varepsilon)|}+\cfrac{\sum_j\left[\pi_-(\varepsilon)\mathcal{D}(\varepsilon)\pi_-^T(\varepsilon)\right]_{ij}}{|\lambda_-(\varepsilon)|} \;+  \\ \cfrac{2\sum_j\left[\pi_+(\varepsilon)\mathcal{D}(\varepsilon)\pi_-^T(\varepsilon)\right]_{ij}}{|\lambda_+(\varepsilon)+\lambda_-(\varepsilon)|} +\cfrac{2\sum_j\left[\pi_+(\varepsilon)\mathcal{D}(\varepsilon)\pi_+^T(\varepsilon)\right]_{ij}}{|\lambda_+(\varepsilon)+\lambda_-(\varepsilon)|}.
\end{multline} 
To quantify the accuracy of the variance, $z_{ii}$, obtained from the projected LNA \eqref{SDE0P}, consider the following limit
\begin{multline}
\lim_{\varepsilon \to 0} \cfrac{\lambda_+(\varepsilon)}{\lambda_-(\varepsilon)}\cdot \cfrac{\sum_j\left[\pi_-(\varepsilon)\mathcal{D}(\varepsilon)\pi_-^T(\varepsilon)\right]_{ij}}{\sum_j\left[\pi_+(\varepsilon)\mathcal{D}(\varepsilon)\pi_+^T(\varepsilon)\right]_{ij}} \\ + \lim_{\varepsilon \to 0}  \cfrac{\lambda_+(\varepsilon)}{\lambda_+(\varepsilon)+\lambda_-(\varepsilon)} \cdot \left[\cfrac{\sum_j\left[\pi_+(\varepsilon)\mathcal{D}(\varepsilon)\pi_-^T(\varepsilon)\right]_{ij}}{\sum_j\left[\pi_+(\varepsilon)\mathcal{D}(\varepsilon)\pi_+^T(\varepsilon)\right]_{ij}} + \cfrac{\sum_j\left[\pi_-(\varepsilon)\mathcal{D}(\varepsilon)\pi_+^T(\varepsilon)\right]_{ij}}{\sum_j\left[\pi_+(\varepsilon)\mathcal{D}(\varepsilon)\pi_+^T(\varepsilon)\right]_{ij}}\right].
\end{multline}
As long as $\pi_+(0)\mathcal{D}_0=\pi_0\mathcal{D}_0=0$, the limit of the bracketed term vanishes as $\varepsilon \to 0$ and $\lambda_+(\varepsilon)\to 0$ (see Appendix, Proposition \ref{prop4}). However, 
\begin{equation}
\lim_{\varepsilon \to 0} \delta(\varepsilon)=:\lim_{\varepsilon \to 0} \cfrac{\lambda_+(\varepsilon)}{\lambda_-(\varepsilon)}\cdot \cfrac{\sum_j\left[\pi_-(\varepsilon)\mathcal{D}(\varepsilon)\pi_-^T(\varepsilon)\right]_{ij}}{\sum_j\left[\pi_+(\varepsilon)\mathcal{D}(\varepsilon)\pi_+^T(\varepsilon)\right]_{ij}} = \cfrac{\lambda_+^{(1)}}{\lambda_-^{(0)}}\cdot \cfrac{\sum_j \mathcal{D}_{0,ij}}{\sum_j \left[\pi_0\mathcal{D}_1\pi_0^T\right]_{ij}}
\end{equation}
will only vanish if the diffusion of the $y_i$ component is limited to the slow timescale (the slow variable, $y_1$, in the standard form \eqref{stochSF} certainly adheres to this requirement). Hence,
\begin{equation}
\lim_{\varepsilon \to 0} \cfrac{\lambda_+(\varepsilon)}{\lambda_-(\varepsilon)} =0 \quad \text{does not imply} \quad \lim_{\varepsilon  \to 0}\delta(\varepsilon) =0,
\end{equation}
which is significant since the difference between $Z^{\infty}_{0,ii}$, the $ii$-th component of the limiting steady-state covariance, and the $ii$-th component of steady-state projected covariance, $Z_{p,ii}^{\infty}$, is determined by $\delta$,
\begin{equation}
Z_{0,ii}^{\infty} = (1+\delta_0)Z_{p,ii}^{\infty},\quad \lim_{\varepsilon \to 0}\delta(\varepsilon)=:\delta_0
\end{equation}
which is the ratio of the fast and slow contributions to the variance as $\varepsilon \to 0$. When $\delta_0 \neq 0$, the projected LNA \eqref{SDE0P} will underestimate the variance of the LNA by a factor of $1+\delta_0$. However, it is sometimes possible to find a coordinate transformation for which the variance of the transformed projected LNA agrees with the covariance of the transformed LNA; see {{\sc figure}} \ref{FIG4} as an example.

In conclusion, this work represents a necessary step towards understanding model reduction methods for stochastic chemical reaction networks, but several open questions remain. First, we did not consider the case in which {\it both} drift eigenvalues vanish in the singular limit. In such cases, the critical manifold will fail to be normally hyperbolic, but this does necessarily limit the applicability of singular perturbation methods in the deterministic realm; see ~\cite{Krupa2001,SzmolyanFoldCanard} and~\citet{kuehn2015}, Chapter 7. The use of more recent techniques to reduce specific LNAs arising from biochemistry in the absence of normal hyperbolicity has not been extensively investigated, although several works have extended deterministic results to singularly perturbed SDEs~\cite{KUEHNstoch,BergPitch,FrenchDucks}. 

Second, as mentioned previously, the LNA is used as a test probe to determine the accuracy of heuristically reduced CMEs (see~\citet{Thomas2012},~\citet{Thomas2011} and \citet{Janssen1989} as examples). For example -- and without submitting too many details -- under steady-state conditions the LNA corresponding to the heuristically-reduced CME of the open Michaelis-Menten network with small enzyme is
\begin{equation*}
{\rm d}y_s = -\cfrac{k_2e_0K_M}{(s+K_M)^2}\cdot y_s \;{\rm d}\tau  +\sqrt{k_0}\;{\rm d}W_1(\tau)-\sqrt{\cfrac{k_2e_0s}{s+K_M}}\;{\rm d}W_2(\tau),
\end{equation*}
whereas the reduction of the LNA is
\begin{equation*}
{\rm d}y_s = -\cfrac{k_2e_0K_M}{(s+K_M)^2}\cdot y_s \;{\rm d}\tau  +\sqrt{k_0}\;{\rm d}W_1(\tau)-\sqrt{\cfrac{k_2e_0s}{s+K_M}\left(1-\cfrac{2K\cdot s}{(s+K_M)^2}\right)}\;{\rm d}W_2(\tau),
\end{equation*}
which led~\citet{Thomas2011} to (correctly) conclude that the heuristic reduction presented by~\citet{Sanft} and~\citet{Rao2003} will not accurately approximate substrate variance unless
\begin{equation}
\eta(s) =: \cfrac{2K\cdot s}{(s+K_M)^2} \ll 1.
\end{equation}
Furthermore, the term $\eta(s)$ is maximal when $s=K_M$, and~\citet{Thomas2011} found that the heuristically reduced CME significantly underestimates the steady-state substrate variance when $s \approx K_M$. But this raises the following question: {\it What does $\eta(s)$ represent, and why does the heuristic reduction of the CME fail {\it only} when $s$ is of the same order of magnitude as $K_M$?} 

The answers to these questions can be found through a careful understanding of how reduction methods work in the presence of eigenvalue disparity and intrinsic noise. We will extend the foundational understanding developed here to address these open questions in forthcoming work(s). 

\section*{Declarations}

\noindent
\textbf{Funding.} Partial funding for this work was provided by the Natural Science and Engineering Research Council of Canada (W.S., RGPIN-2021-02747). \\
\noindent
\textbf{Competing Interests.} The authors have no relevant financial or non-financial interests to disclose. \\
\noindent
\textbf{Data Availability.} The authors declare that the data supporting the findings of this manuscript are available within the paper.

%\section*{Compliance with Ethical Standards}

\section*{Classification}

\noindent
\textbf{Mathematics Subject Classification:} 34D15, 34E15, 60J70, 92C45 and 92E20

\newpage
\section*{Appendix}

This appendix contains three subsections. In subsection \eqref{subA} we recall some basic facts about matrices with simple eigenvalues that depend on a parameter, $\varepsilon$, that are analytic in a neighborhood of $\varepsilon$. The details of these facts can be found in~\citet{greenbaum2019}. In subsection \ref{subB}, we prove that the long-time covariance, $Z_{\infty}$ converges to $Z^{\infty}_0$ as $\varepsilon \to 0$, provided the blanket assumptions discussed in subsection \ref{subA}, as well as the Fredholm alternative \eqref{FAL}, hold. In subsection \ref{subC} we prove that the norm of the steady-state covariance, $||Z^{\infty}||$, is unbounded as $\varepsilon \to 0$ if the Fredholm alternative fails to hold. 

\subsection{Blanket assumptions: First-order perturbation theory for simple eigenvalues}\label{subA}

We will assume that $A=A_0+\varepsilon A_1$ has two distinct, strictly negative eigenvalues, $\lambda_{\pm}$ with corresponding eigenvectors $v_{\pm}$. Moreover, we will assume that $A_0$ is singular, with one strictly negative eigenvalue and one eigenvalue that is identically zero. Since the eigenvalues of $A_0$ are simple, there exists a projection matrix, $\pi_0$, that projects onto $\ker A_0$ along the direction of $A_0$'s image:
\begin{subequations}
\begin{align}
\pi_0&:\mathbb{R}^2 \to \ker A_0,\\
I-\pi_0&:\mathbb{R}^2 \to {\rm image} \;A_0.
\end{align}
\end{subequations}

As long as $A(\varepsilon)$ is analytic in a neighborhood of $\varepsilon =0$, then $A(\varepsilon)$ has eigenvalues, $\lambda_{\pm}(\varepsilon)$ with $\lambda_-\ll \lambda_+ < 0$, that are analytic in a neighborhood of $\varepsilon =0$ and 
\begin{subequations}
\begin{align}
\lambda_+(\varepsilon) &= \varepsilon \cdot \cfrac{{\rm d} \lambda_+(\varepsilon)}{{\rm d}\varepsilon }\bigg|_{\varepsilon =0} +\mathcal{O}(\varepsilon^2) =:\varepsilon\lambda_+^{(1)}+\mathcal{O}(\varepsilon^2),\\
\lambda_-(\varepsilon) &= \lambda_-(0)+\varepsilon \cdot \cfrac{{\rm d} \lambda_-(\varepsilon)}{{\rm d}\varepsilon }\bigg|_{\varepsilon =0} +\mathcal{O}(\varepsilon^2) =:\lambda_-^{(0)}+\varepsilon\lambda_-^{(1)}+\mathcal{O}(\varepsilon^2)
\end{align}
\end{subequations}
Moreover, the first-order corrections $\lambda_+^{(0)}$ and $\lambda_-^{(1)}$ are
\begin{equation}\label{eigs}
\lambda_+^{(1)} = \cfrac{w^TA_1v_+^{(0)}}{w^Tv_+^{(0)}} = {\rm trace}(\pi_0A_1), \quad \lambda_-^{(1)} = {\rm trace}\left((I-\pi_0)A_1\right),
\end{equation}
where $w^TA_0=0$, $A_0v_+^{(0)}=0$, and $A_0v_-^{(0)}=\lambda_-^{(0)}v_-^{(0)}.$ Note that it follows from \eqref{eigs} that
\begin{equation}
\pi_0A_1\pi_0v = \lambda_+^{(1)}\pi_0v, \quad \forall v\in \mathbb{R}^2,\quad \text{since}\quad \pi_0 = \cfrac{v_+^{(0)}w^T}{w^Tv_+^{(0)}}= I - \cfrac{1}{\lambda_-^{(0)}}A_0
\end{equation}

The eigenvectors of $A(\varepsilon)$, $v_{\pm}(\varepsilon)$, can also be expanded to first order, but we will not present these expansions here. The important component of our analysis pertains to the expansion of the projection operators, $\pi_{\pm}(\varepsilon)$,
\begin{subequations}
\begin{align}
\pi_+(\varepsilon)&:\mathbb{R}^2 \to {\rm span}\{v_+\},\\
I-\pi_+(\varepsilon):=\pi_-(\varepsilon)&:\mathbb{R}^2 \to {\rm span}\{v_-\},
\end{align}
\end{subequations}
that are analytic in a neighborhood of $\varepsilon =0$. The leading order terms in the expansion are $\pi_+(0)=\pi_0$ and $\pi_-(0)=I-\pi_0$, with the first order corrections given by
\begin{subequations}
\begin{align}
\pi_+^{(1)} &= -\pi_0A_1(I-\pi_0)-(I-\pi_0)A_1\pi_0,\\
\pi_-^{(1)} &= -(I-\pi_0)A_1\pi_0 - \pi_0A_1(I-\pi_0).
\end{align}
\end{subequations}
Thus, we have
\begin{subequations}\label{expans0}
\begin{align}
\pi_+ &= \quad \;\;\;\pi_0 + \varepsilon \pi_+^{(1)}  + \mathcal{O}(\varepsilon^2),\\
\pi_- &= I-\pi_0 + \varepsilon \pi_-^{(1)} + \mathcal{O}(\varepsilon^2).
\end{align}
\end{subequations}

\subsection{Convergence of the steady-state covariance, $Z_{\infty}$ as $\varepsilon \to 0$}\label{subB}
Let $A(\varepsilon)=A_0+\varepsilon A_1$ have two distinct eigenvalues, $\lambda_{\pm}$ and corresponding eigenvectors $v_{\pm}$. Then, the Lyapunov operator,
\begin{equation}
\mathcal{L}(Z) = (A_0+\varepsilon A_1)Z+ Z(A_0+\varepsilon A_1)^T,
\end{equation}
has the corresponding eigenvalue/eigenvector pairs:
\begin{equation}
(2\lambda_+,v_+v_+^T),\quad (\lambda_-+\lambda_+,v_-v_+^T),\quad (\lambda_-+\lambda_+,v_+v_-^T),\quad (2\lambda_-,v_-v_-^T).
\end{equation}

Let $\pi_+$ denote the projection matrix that projects onto ${\rm span}\{v_+\}$ along $v_-$, and let $\pi_-$ project onto ${\rm span}\{v_-\}$ along the direction of $v_+$ with the identities
\begin{subequations}\label{expans}
\begin{align*}
\pi_+ &= \quad \;\;\;\pi_0 + \varepsilon \pi_+^{(1)}  + \mathcal{O}(\varepsilon^2),\\
\pi_- &= I-\pi_0 + \varepsilon \pi_-^{(1)} + \mathcal{O}(\varepsilon^2),
\end{align*}
\end{subequations}
from \eqref{expans0}. The steady-state covariance is the solution to the Lyapunov equation
\begin{equation}
\mathcal{L}(Z) = -B_0B_0^T - \varepsilon B_1B_1^T,
\end{equation}
which can be solved in stages thanks to its linearity.

\begin{proposition}\label{prop4}
The steady-state covariance, $Z^{\infty}$, converges to 
\begin{equation*}
\lim_{\varepsilon \to 0} Z^{\infty} = Z^{\infty}_0 = \cfrac{B_0B_0^T}{2|\lambda_-^{(0)}|} + \cfrac{\pi_0B_1B_1^T\pi_0^T}{2| \lambda_+^{(1)}|}
\end{equation*}
as $\varepsilon \to 0$ if the Fredholm alternative holds and $\pi_0B_0 = 0$.
\end{proposition}

\begin{proof}
First, project the right-hand-side of the Lyapunov equation onto its respective eigenspaces:
\begin{equation}
-\mathcal{L}(Z_0) = \pi_-B_0B_0^T\pi_-^T +\pi_-B_0B_0^T\pi_+ +  \pi_+B_0B_0^T\pi_-^T + \pi_+B_0B_0^T\pi_+^T.
\end{equation}
The {\it action} of inverting $\mathcal{L}$ yields
\begin{equation}\label{part1}
Z_0 = \cfrac{\pi_-B_0B_0^T\pi_-^T}{2|\lambda_-|} +\cfrac{\pi_-B_0B_0^T\pi_+}{|\lambda_-+\lambda_+|} +  \cfrac{\pi_+B_0B_0^T\pi_-^T}{|\lambda_-+\lambda_+|} + \cfrac{\pi_+B_0B_0^T\pi_+^T}{2|\lambda_+|}.
\end{equation}
Next, we can expand each term on the right-hand side of \eqref{part1} in terms of $\varepsilon$. Starting with the first term, we have
\begin{equation}
\cfrac{\pi_-B_0B_0^T\pi_-^T}{2|\lambda_-|} = \cfrac{B_0B_0^T -\pi_0B_0B_0^T\pi_0^T +\mathcal{O}(\varepsilon)}{2|\lambda_-^{(0)}+\varepsilon \lambda_-^{(1)} + \mathcal{O}(\varepsilon^2)|} 
\end{equation}
which reduces to
\begin{equation}\label{part1A}
\cfrac{\pi_-B_0B_0^T\pi_-^T}{2|\lambda_-|} = \cfrac{B_0B_0^T + \mathcal{O}(\varepsilon)}{2|\lambda_-^{(0)}+\varepsilon \lambda_-^{(1)} + \mathcal{O}(\varepsilon^2)|} 
\end{equation}
since $\pi_0B_0=0$ by Proposition 1. Taking the limit as $\varepsilon \to 0$ yields
\begin{equation}
\lim_{\varepsilon \to 0}\left(\cfrac{B_0B_0^T + \mathcal{O}(\varepsilon)}{2|\lambda_-^{(0)}+\varepsilon \lambda_-^{(1)} + \mathcal{O}(\varepsilon^2)|}\right) = \cfrac{B_0B_0^T}{2|\lambda_-^{(0)}|} .
\end{equation}
A straightforward calculation reveals the middle two terms on the right-hand-side of \eqref{part1} vanish as $\varepsilon \to 0$. This leaves the last term,
\begin{equation}
\cfrac{\pi_+B_0B_0^T\pi_+^T}{2|\lambda_+|} = \cfrac{\pi_0B_0B_0^T\pi_0^T + \varepsilon \pi_+^{(1)}B_0B_0^T\pi_0 + \varepsilon \pi_0B_0B_0^T\pi_+^{(1)} + \varepsilon^2 \pi_+^{(1)}B_0B_0^T\pi_+^{(1),T}}{2|\varepsilon \lambda_+^{(1)}+\mathcal{O}(\varepsilon^2)|},
\end{equation}
which (again, due to Proposition 1) reduces to 
\begin{equation}
\cfrac{\pi_+B_0B_0^T\pi_+^T}{2|\lambda_+|} = \cfrac{\varepsilon^2 \pi_+^{(1)}B_0B_0^T\pi_+^{(1),T}}{2|\varepsilon \lambda_+^{(1)}+\mathcal{O}(\varepsilon^2)|},
\end{equation}
and vanishes in the limit as $\varepsilon \to 0.$

Now consider the second stage,
\begin{equation}
-\mathcal{L}(Z_1) = \varepsilon\left(\pi_-B_1B_1^T\pi_-^T +\pi_-B_1B_1^T\pi_+ +  \pi_+B_1B_1^T\pi_-^T + \pi_+B_1B_1^T\pi_+^T\right).
\end{equation}
Again, the action of inverting $\mathcal{L}$ yields
\begin{equation}\label{part2}
Z_1 = \varepsilon \cdot \left(\cfrac{\pi_-B_1B_1^T\pi_-^T}{2|\lambda_-|} +\cfrac{\pi_-B_1B_1^T\pi_+}{|\lambda_-+\lambda_+|} +  \cfrac{\pi_+B_1B_1^T\pi_-^T}{|\lambda_-+\lambda_+|} + \cfrac{\pi_+B_1B_1^T\pi_+^T}{2|\lambda_+|}\right).
\end{equation}
The first 3 terms on the right-hand-side of \eqref{part2} vanish as $\varepsilon \to 0$. This leaves only the last term,
\begin{equation}
\varepsilon \cdot \cfrac{\pi_+B_1B_1^T\pi_+^T}{2|\lambda_+|} = \cfrac{\varepsilon \pi_0B_1B_1^T\pi_0^T + \varepsilon^2 \pi_+^{(1)}B_1B_1^T\pi_0 + \varepsilon^2 \pi_0B_1B_1^T\pi_+^{(1)} + \varepsilon^3 \pi_+^{(1)}B_1B_1^T\pi_+^{(1),T}}{2|\varepsilon \lambda_+^{(1)}+\mathcal{O}(\varepsilon^2)|},
\end{equation}
which converges to
\begin{equation}
\lim_{\varepsilon \to 0 }\left(\cfrac{\varepsilon \pi_0B_1B_1^T\pi_0^T}{2|\varepsilon \lambda_+^{(1)}+\mathcal{O}(\varepsilon^2)|}\right) = \cfrac{\pi_0B_1B_1^T\pi_0^T}{2| \lambda_+^{(1)}|}.
\end{equation}
Summing $Z_0$ and $Z_1$ yields the steady-state covariance in the limit as $\varepsilon \to 0$:
\begin{equation}
\lim_{\varepsilon \to 0} Z^{\infty} = Z^{\infty}_0 = \cfrac{B_0B_0^T}{2|\lambda_-^{(0)}|} + \cfrac{\pi_0B_1B_1^T\pi_0^T}{2| \lambda_+^{(1)}|}.
\end{equation}
\end{proof}

\subsection{The Fredholm alternative and unbounded steady-state covariance}\label{subC}

\begin{proposition}\label{prop5}
The steady-state covariance, $Z^{\infty}$, satisfies
\begin{equation}\label{LA}
\mathcal{L}_0(Z)+ \varepsilon \mathcal{L}_1(Z) = -B_0B_0^T -\varepsilon B_1B_1^T,
\end{equation}
with $\mathcal{L}_0(Z)=A_0Z+ZA_0^T$ and $\mathcal{L}_1(Z)=A_1Z+ZA_1^T$. Suppose the Fredholm alternative \eqref{FAL} fails to hold and 
\begin{equation}
\pi_0B_0 \neq 0.
\end{equation}
Then, the steady-state covariance, $Z_{\infty}$, is unbounded as $\varepsilon \to 0$:
\begin{equation}
||Z^{\infty}|| \to \infty\;\;\text{as} \;\;\varepsilon \to 0.
\end{equation}
\end{proposition}
\begin{proof}
Project the right-hand-side of \eqref{LA} onto the respective eigenspaces of $\mathcal{L}$ and compute the action of $\mathcal{L}^{-1}$ by dividing each projected term by its corresponding eigenvalue. The projection onto the slow eigenspace, $v_+v_+^T$, is
\begin{equation}\label{expansA}
\cfrac{\pi_+B_0B_0^T\pi_+^T}{2|\lambda_+|} = \cfrac{\pi_0B_0B_0^T\pi_0^T + \varepsilon \pi_+^{(1)}B_0B_0^T\pi_0 + \varepsilon \pi_0B_0B_0^T\pi_+^{(1),T} + \varepsilon^2 \pi_+^{(1)}B_0B_0^T\pi_+^{(1),T}}{2|\varepsilon \lambda_+^{(1)}+\mathcal{O}(\varepsilon^2)|}.
\end{equation}
However, if we examine the limiting behavior as $\varepsilon \to 0$, the first term on the right-hand-side of \eqref{expansA} has no limit as $\varepsilon \to 0$ since $\pi_0B_0 \neq 0$,
\begin{equation}
\lim_{\varepsilon \to 0} \cfrac{\pi_0B_0B_0^T\pi_0^T}{2|\varepsilon \lambda_+^{(1)}+\mathcal{O}(\varepsilon^2)|}\quad \text{does not exist},
\end{equation}
and the assertion follows.
\end{proof}
\newpage
\bibliographystyle{elsarticle-num-names}
\bibliography{Eigenvalues.bib}

@article {HekGSPT,
    AUTHOR = {Hek, Geertje},
     TITLE = {Geometric singular perturbation theory in biological practice},
   JOURNAL = {J. Math. Biol.},
  FJOURNAL = {Journal of Mathematical Biology},
    VOLUME = {60},
      YEAR = {2010},
    NUMBER = {3},
     PAGES = {347--386},
}

@article{BurrageCLE,
    author = {Mélykúti, Bence and Burrage, Kevin and Zygalakis, Konstantinos C.},
    title = {Fast stochastic simulation of biochemical reaction systems by alternative formulations of the chemical Langevin equation},
    journal = {The Journal of Chemical Physics},
    volume = {132},
    number = {16},
    pages = {164109},
    year = {2010},
}

@article{SzmolyanFoldCanard,
author = {Krupa, M. and Szmolyan, P.},
title = {Extending Geometric Singular Perturbation Theory to Nonhyperbolic Points---Fold and Canard Points in Two Dimensions},
journal = {SIAM Journal on Mathematical Analysis},
volume = {33},
number = {2},
pages = {286-314},
year = {2001}
}

@article{GrimaSecondOrder,
  title = {Linear-noise approximation and the chemical master equation agree up to second-order moments for a class of chemical systems},
  author = {Grima, Ramon},
  journal = {Phys. Rev. E},
  volume = {92},
  issue = {4},
  pages = {042124},
  numpages = {10},
  year = {2015},
}

@article {FrenchDucks,
    AUTHOR = {Berglund, Nils and Gentz, Barbara and Kuehn, Christian},
     TITLE = {Hunting {F}rench ducks in a noisy environment},
   JOURNAL = {J. Differential Equations},
  FJOURNAL = {Journal of Differential Equations},
    VOLUME = {252},
      YEAR = {2012},
    NUMBER = {9},
     PAGES = {4786--4841},
}

@article {BergPitch,
    AUTHOR = {Berglund, Nils and Gentz, Barbara},
     TITLE = {Pathwise description of dynamic pitchfork bifurcations with
              additive noise},
   JOURNAL = {Probab. Theory Related Fields},
  FJOURNAL = {Probability Theory and Related Fields},
    VOLUME = {122},
      YEAR = {2002},
    NUMBER = {3},
     PAGES = {341--388},
}

@article{Unreasonable,
title = {The unreasonable effectiveness of the total quasi-steady state approximation, and its limitations},
journal = {Journal of Theoretical Biology},
volume = {583},
pages = {111770},
year = {2024},
author = {Justin Eilertsen and Santiago Schnell and Sebastian Walcher},
}

@misc{greenbaum2019,
      title={First-order Perturbation Theory for Eigenvalues and Eigenvectors}, 
      author={Anne Greenbaum and Ren-cang Li and Michael L. Overton},
      year={2019},
      eprint={1903.00785},
      archivePrefix={arXiv},
      primaryClass={math.NA},
      url={https://arxiv.org/abs/1903.00785}, 
}

@article{ganguly2025asymptotic,
  title={Asymptotic Analysis of the Total Quasi-Steady State Approximation for the {M}ichaelis--{M}enten Enzyme Kinetic Reactions},
  author={Ganguly, Arnab and KhudaBukhsh, Wasiur R},
  journal={arXiv preprint arXiv:2503.20145},
  year={2025},
}

@article{BerglundGentzJDE,
title = {Geometric singular perturbation theory for stochastic differential equations},
journal = {Journal of Differential Equations},
volume = {191},
number = {1},
pages = {1-54},
year = {2003},
author = {Nils Berglund and Barbara Gentz}
}

@article {KnoblochCM,
    AUTHOR = {Knobloch, E. and Wiesenfeld, K. A.},
     TITLE = {Bifurcations in fluctuating systems: the center-manifold
              approach},
   JOURNAL = {J. Statist. Phys.},
  FJOURNAL = {Journal of Statistical Physics},
    VOLUME = {33},
      YEAR = {1983},
    NUMBER = {3},
     PAGES = {611--637},
}

@article{KUEHNstoch,
title = {A mathematical framework for critical transitions: Bifurcations, fast–slow systems and stochastic dynamics},
journal = {Physica D: Nonlinear Phenomena},
volume = {240},
number = {12},
pages = {1020-1035},
year = {2011},
author = {Christian Kuehn}
}

@article {kurtz1978,
    AUTHOR = {Kurtz, Thomas G.},
     TITLE = {Strong approximation theorems for density dependent {M}arkov
              chains},
   JOURNAL = {Stochastic Process. Appl.},
  FJOURNAL = {Stochastic Processes and their Applications},
    VOLUME = {6},
      YEAR = {1977/78},
    NUMBER = {3},
     PAGES = {223--240}
}

@article{Yin2005,
title = {Limit behavior of two-time-scale diffusions revisited},
journal = {Journal of Differential Equations},
volume = {212},
number = {1},
pages = {85-113},
year = {2005},
author = {R.Z. Khasminskii and G. Yin}
}

@article{Yin2004,
author = {Khasminskii, R. Z. and Yin, G.},
title = {On Averaging Principles: An Asymptotic Expansion Approach},
journal = {SIAM Journal on Mathematical Analysis},
volume = {35},
number = {6},
pages = {1534-1560},
year = {2004}
}

@article {Kang2019,
    AUTHOR = {Kang, Hye-Won and KhudaBukhsh, Wasiur R. and Koeppl, Heinz and
              Rempa{\l}a, Grzegorz A.},
     TITLE = {Quasi-steady-state approximations derived from the stochastic
              model of enzyme kinetics},
   JOURNAL = {Bull. Math. Biol.},
  FJOURNAL = {Bulletin of Mathematical Biology. A Journal Devoted to
              Research at the Interface of the Life and Mathematical
              Sciences},
    VOLUME = {81},
      YEAR = {2019},
    NUMBER = {5},
     PAGES = {1303--1336}
}

@article{Krupa2001,
title={Extending slow manifolds near transcritical and pitchfork singularities},
journal = {Nonlinearity},
fjournal = {Nonlinearity},
Volume = {14},
year ={2001},
pages ={1473--1491},
author ={M. Krupa and P. Szmolyan}
}

@article{Janssen1989,
title = {The elimination of fast variables in complex chemical reactions. {I}{I}{I}. Mesoscopic level},
fjournal = {Journal of Statistical Physics},
journal = {J. Stat. Phys.},
volume = {57},
pages = {187-198},
year = {1989},
author = {Janssen, J. A. M.}
}

@article{Burrage2008,
author = {MacNamara,Shev  and Bersani,Alberto M.  and Burrage,Kevin  and Sidje,Roger B. },
title = {Stochastic chemical kinetics and the total quasi-steady-state assumption: Application to the stochastic simulation algorithm and chemical master equation},
journal = {The Journal of Chemical Physics},
volume = {129},
number = {9},
pages = {095105},
year = {2008}
}

@article{Rao2003,
author = {Rao,Christopher V.  and Arkin,Adam P. },
title = {Stochastic chemical kinetics and the quasi-steady-state assumption: Application to the Gillespie algorithm},
fjournal = {The Journal of Chemical Physics},
journal = {J. Chem. Phys.},
volume = {118},
number = {11},
pages = {4999-5010},
year = {2003}
}

@article{Thomas2012,
title={The slow-scale linear noise approximation: an accurate, reduced stochastic description of biochemical networks under timescale separation conditions},
author={Thomas, Philipp and Straube, Arthur V. and Grima, Ramon},
fjournal={BMC Systems Biology},
journal={BMC Sys. Biol.},
Volume={6},
Number={1},
pages={39},
year={2012}
}

@article{Fenichel1979,
    AUTHOR = {Fenichel, Neil},
     TITLE = {Geometric singular perturbation theory for ordinary
              differential equations},
   JOURNAL = {J. Differ. Equations},
  FJOURNAL = {Journal of Differential Equations},
    VOLUME = {31},
      YEAR = {1979},
     PAGES = {53--98}
}

@book{Berglund,
    AUTHOR = {Berglund, Nils and Gentz, Barbara},
     TITLE = {Noise-induced phenomena in slow-fast dynamical systems},
 PUBLISHER = {Springer-Verlag London, Ltd., London},
      YEAR = {2006},
     PAGES = {xiv+276},
   MRCLASS = {37H99 (34C29 34F05 34K50 37G15 60F10 60H10 82C05)},
  MRNUMBER = {2197663},
MRREVIEWER = {Dirk Bl\"omker}
}

@article{Borghans1996,
title={Extending the quasi-steady state approximation by changing variables},
  author={Borghans, Jos{\'e} A. M. and De Boer, Rob J. and Segel, Lee A.},
  fjournal={Bulletin of Mathematical Biology},
  journal = {Bull. Math. Biol.},
  volume={58},
  pages={43--63},
  year={1996},
  publisher={Springer}
}

@article{Goeke2015,
    AUTHOR = {Goeke, Alexandra and Walcher, Sebastian and Zerz, Eva},
     TITLE = {Determining ``small parameters'' for quasi-steady state},
   JOURNAL = {J. Differ. Equations.},
  FJOURNAL = {Journal of Differential Equations},
    VOLUME = {259},
      YEAR = {2015},
     PAGES = {1149--1180}
}

@book {kuehn2015,
    AUTHOR = {Kuehn, Christian},
     TITLE = {Multiple time scale dynamics},
    SERIES = {Applied Mathematical Sciences},
    VOLUME = {191},
 PUBLISHER = {Springer},
      YEAR = {2015},
     PAGES = {xiv+814},
   MRCLASS = {34-02 (34-01 34Exx 37-02 37C10 37Gxx)},
  MRNUMBER = {3309627},
MRREVIEWER = {Tewfik Sari}
}

@book{Wechselberger2020,
    AUTHOR = {Wechselberger, Martin},
     TITLE = {Geometric Singular Perturbation Theory Beyond the Standard Forms},
     SERIES = {Frontiers in Applied dynamical systems: Tutorials and Reviews},
     NUMBER = {6},
 PUBLISHER = {Springer},
      YEAR = {2020}
}

@article{Kim2020,
    author = {Kim, Jae Kyoung AND Tyson, John J.},
    fjournal = {PLoS Computational Biology},
    Journal = {PLoS Comp. Biol.},
    publisher = {Public Library of Science},
    title = {Misuse of the {M}ichaelis–-{M}enten rate law for protein interaction networks and its remedy},
    year = {2020},
    month = {10},
    volume = {16},
    pages = {1-21},
    number = {10}
}

@article{Stroberg2016,
author = {Stroberg, Wylie and Schnell, Santiago},
fjournal = {Biophysical Chemistry},
journal = {Biophys. Chem.},
pages = {17--27},
title = {On the estimation errors of ${K}_{M}$ and $v$ from time-course experiments using the {M}ichaelis–{M}enten equation},
volume = {219},
year = {2016}
}

@article{Agarwal2012,
author = {Agarwal,Animesh  and Adams,Rhys  and Castellani,Gastone C.  and Shouval,Harel Z. },
title = {On the precision of quasi steady state assumptions in stochastic dynamics},
journal = {The Journal of Chemical Physics},
volume = {137},
number = {4},
pages = {044105},
year = {2012}
}

@article{Thomas2011,
author = {Thomas, Philipp and Straube, Arthur V. and Grima, Ramon },
title = {Communication: Limitations of the stochastic quasi-steady-state approximation in open biochemical reaction networks},
fjournal = {The Journal of Chemical Physics},
journal = {J. Chem. Phys.},
volume = {135},
number = {18},
pages = {181103},
year = {2011}
}

@article{Herath,
author = {Herath, Narmada  and Del Vecchio,Domitilla },
title = {Reduced linear noise approximation for biochemical reaction networks with time-scale separation: The stochastic t{Q}{S}{S}{A}$^+$},
journal = {J. Chem. Phys.},
fjournal = {The Journal of Chemical Physics},
volume = {148},
number = {9},
pages = {094108},
year = {2018}
}

@article{JSEssLNA,
	Author = {Eilertsen, Justin and Srivastava, Kashvi and Schnell, Santiago},
	Journal = {Journal of Mathematical Biology},
	Number = {1},
	Pages = {3},
	Title = {Stochastic enzyme kinetics and the quasi-steady-state reductions: Application of the slow scale linear noise approximation {\`a} la Fenichel},
	Volume = {85},
	Year = {2022},
 }

@article{ThomasPO,
  title = {Rigorous elimination of fast stochastic variables from the linear noise approximation using projection operators},
  author = {Thomas, Philipp and Grima, Ramon and Straube, Arthur V.},
  journal = {Phys. Rev. E},
  volume = {86},
  issue = {4},
  pages = {041110},
  numpages = {9},
  year = {2012}
}

@article{KIM2015,
author = {Kim, Jae Kyoung and Josić, Krešimir and Bennett, Matthew R.},
title = {The relationship between stochastic and deterministic quasi-steady state approximations},
fjournal = {BMC Systems Biology},
journal = {BMC Syst. Biol.},
volume = {9},
pages = {87},
year = {2015}
}

@article{Mastny2007,
author = {Mastny, Ethan A.  and Haseltine, Eric L.  and Rawlings, James B. },
title = {Two classes of quasi-steady-state model reductions for stochastic kinetics},
fjournal = {The Journal of Chemical Physics},
journal = {J. Chem. Phys.},
volume = {127},
number = {9},
pages = {094106},
year = {2007}
}

@article{Cao,
author = {Cao,Yang  and Gillespie,D. T.  and Petzold, L. R. },
title = {The slow-scale stochastic simulation algorithm},
fjournal = {The Journal of Chemical Physics},
journal = {J. Chem. Phys.},
volume = {122},
number = {1},
pages = {014116},
year = {2005}
}

@article{KIM2014,
title = {The Validity of Quasi-Steady-State Approximations in Discrete Stochastic Simulations},
fjournal = {Biophysical Journal},
journal = {Biophys. J.},
volume = {107},
number = {3},
pages = {783 - 793},
year = {2014},
author = {Jae Kyoung Kim and Krešimir Josić and Matthew R. Bennett}
}

@article{GILLESPIECME,
title = {A rigorous derivation of the chemical master equation},
fjournal = {Physica A: Statistical Mechanics and its Applications},
journal = {Physica A},
author = {Gillespie, D. T.},
volume = {188},
pages = {404--425},
year = {1992}
}

@article{Sanft,
author = {Sanft, K.R.  and Gillespie, D. T.  and Petzold, L. R. },
title = {The legitimacy of the stochastic {M}ichaelis--{M}enten approximation},
fjournal = {IET Systems Biology},
journal = {IET Syst. Biol.},
volume = {5},
pages = {58--69},
year = {2011},
}

@article {OpenMMin,
author = {Eilertsen, Justin and Roussel, Marc and Schnell, Santiago and Walcher, Sebastian},
title = {On the quasi-steady-state approximation in an open {M}ichaelis--{M}enten reaction mechanism},
fjournal = {{AIMS} Mathematics},
journal = {{AIMS} Math},
issn = {2473-6988},
year ={2021},
volume = {6},
issue={7},
pages = {6781–-6814}
}

@article{PAHLAJANI201196,
title = {Stochastic reduction method for biological chemical kinetics using time-scale separation},
fjournal = {Journal of Theoretical Biology},
journal = {J. Theo. Biol.},
volume = {272},
number = {1},
pages = {96-112},
year = {2011},
author = {Chetan D. Pahlajani and Paul J. Atzberger and Mustafa Khammash}
}

@article {Roberts2015,
    AUTHOR = {Chen, Xiaopeng and Roberts, Anthony J. and Duan, Jinqiao},
     TITLE = {Centre manifolds for stochastic evolution equations},
   JOURNAL = {J. Difference Equ. Appl.},
  FJOURNAL = {Journal of Difference Equations and Applications},
    VOLUME = {21},
      YEAR = {2015},
    NUMBER = {7},
     PAGES = {606--632}
}

@article{GrimaBreakdown,
  title = {Noise-Induced Breakdown of the Michaelis-Menten Equation in Steady-State Conditions},
  author = {Grima, R.},
  journal = {Phys. Rev. Lett.},
  volume = {102},
  issue = {21},
  pages = {218103},
  numpages = {4},
  year = {2009},
}

@article {Roberts1996,
    AUTHOR = {Xu, Chao and Roberts, A. J.},
     TITLE = {On the low-dimensional modelling of {S}tratonovich stochastic
              differential equations},
   JOURNAL = {Phys. A},
  FJOURNAL = {Physica A},
    VOLUME = {225},
      YEAR = {1996},
    NUMBER = {1},
     PAGES = {62--80}
}

@article {Roberts2013,
    AUTHOR = {Wang, W. and Roberts, A. J.},
     TITLE = {Slow manifold and averaging for slow-fast stochastic
              differential system},
   JOURNAL = {J. Math. Anal. Appl.},
  FJOURNAL = {Journal of Mathematical Analysis and Applications},
    VOLUME = {398},
      YEAR = {2013},
    NUMBER = {2},
     PAGES = {822--839}
}

@article{Kumar1998,
title = {Singular perturbation modeling of nonlinear processes with nonexplicit time-scale multiplicity},
journal = {Chemical Engineering Science},
volume = {53},
number = {8},
pages = {1491-1504},
year = {1998},
author = {Aditya Kumar and Panagiotis D. Christofides and Prodromos Daoutidis}
}

@article {Wilhelm2000,
    AUTHOR = {Schneider, Klaus R. and Wilhelm, Thomas},
     TITLE = {Model reduction by extended quasi-steady-state approximation},
   JOURNAL = {J. Math. Biol.},
  FJOURNAL = {Journal of Mathematical Biology},
    VOLUME = {40},
      YEAR = {2000},
    NUMBER = {5},
     PAGES = {443--450}
}

@article {OthmerI,
    AUTHOR = {Lee, Chang Hyeong and Othmer, Hans G.},
     TITLE = {A multi-time-scale analysis of chemical reaction networks.
              {I}. {D}eterministic systems},
   JOURNAL = {J. Math. Biol.},
  FJOURNAL = {Journal of Mathematical Biology},
    VOLUME = {60},
      YEAR = {2010},
    NUMBER = {3},
     PAGES = {387--450}
}

@article {OthmerII,
    AUTHOR = {Kan, Xingye and Lee, Chang Hyeong and Othmer, Hans G.},
     TITLE = {A multi-time-scale analysis of chemical reaction networks:
              {II}. {S}tochastic systems},
   JOURNAL = {J. Math. Biol.},
  FJOURNAL = {Journal of Mathematical Biology},
    VOLUME = {73},
      YEAR = {2016},
    NUMBER = {5},
     PAGES = {1081--1129}
}

\end{document}